\documentclass[%
reprint, 
superscriptaddress,
amsmath,amssymb,
aps,
prx,
]{revtex4-2}

\usepackage[utf8]{inputenc} 
\usepackage[T1]{fontenc}    
\usepackage[colorlinks=true, allcolors=blue]{hyperref}
\usepackage{url}            
\usepackage{booktabs}       
\usepackage{amsfonts}       
\usepackage{nicefrac}       
\usepackage{microtype}      
\usepackage[dvipsnames]{xcolor}         
\usepackage{graphicx}
\usepackage{amsmath}
\usepackage{physics}
\usepackage{dsfont}
\usepackage{rotating}
\usepackage{multirow}
\usepackage{bigdelim}
\usepackage{algorithm}
\usepackage[noend]{algorithmic}
\usepackage{subcaption}
\usepackage{siunitx}
\usepackage{titletoc}

\newcommand{\fig}{Fig.}

\newcommand{\Fig}{Figure}
\newcommand{\Figs}{Figures}
\newcommand{\tab}{Tab.}

\newcommand{\app}{App.}
\newcommand{\apps}{Apps.}

\newcommand{\seclab}{Sec.}

\newcommand{\eq}{Eq.}
\newcommand{\eqs}{Eqs.}
\newcommand{\Eq}{Eq.}

\newcommand{\reflab}{Ref.}

\newcommand{\spinup}{\lvert\uparrow\rangle}
\newcommand{\spindown}{\lvert\downarrow\rangle}
\newcommand{\dataset}{\mathcal{D}}
\newcommand{\datasize}{N}

\newcommand{\snapshot}{S}
\newcommand{\systemsize}{D}
\newcommand{\transverse}{\Omega}
\newcommand{\longitudinal}{\delta}
\newcommand{\deltaxy}{\longitudinal_{\rm XY}}
\newcommand{\actbottle}{z}

\newcommand{\loss}{L}

\definecolor{green}{rgb}{.2,.6,.2}
\definecolor{brickred}{rgb}{0.8, 0.25, 0.33}
\definecolor{brightcerulean}{rgb}{0.11, 0.67, 0.84}
\definecolor{maroon}{rgb}{0.788, 0.0, 0.086}
\definecolor{ao}{rgb}{0.0, 0.5, 0.0}

\newcommand{\figcap}[1]{\textbf{#1}}
\newcommand{\highlight}[1]{\textit{#1}}

\usepackage{tikz}
\usetikzlibrary{positioning}
\usepackage{mathtools}

\usepackage{xcolor}

\definecolor{MyGreen}{HTML}{7FDF57}
\definecolor{MyBlue}{HTML}{3E52E1}
\definecolor{MyRed}{HTML}{FF4D4D}
\definecolor{White}{HTML}{FFFFFF}

\colorlet{LightGreen}{MyGreen!40}
\colorlet{LightBlue}{MyBlue!40}
\colorlet{LightRed}{MyRed!40}

\colorlet{DarkGreen}{MyGreen!80!black}
\colorlet{DarkBlue}{MyBlue!80!black}
\colorlet{DarkRed}{MyRed!80!black}

\newcommand{\bs}{\blacksquare}
\newcommand{\sq}{\square}

\DeclareRobustCommand{\sus}[1]{\substack{#1}} 
\newcommand{\us}{\,\mu\mathrm{s}}

\begin{document}

\title{\texorpdfstring{TetrisCNN for interpretable detection of phases of matter\\ from experimental quantum simulator data}{TetrisCNN for interpretable detection of phases of matter from experimental quantum simulator data}}

\author{Kacper~Cybiński}
\thanks{These two authors contributed equally.}
\affiliation{Faculty of Physics, University of Warsaw, Pasteura 5, 02-093 Warsaw, Poland}
\affiliation{IDEAS Research Institute, Królewska 27, 00-060 Warsaw, Poland}

\author{Björn van Zwol}
\thanks{These two authors contributed equally.}
\affiliation{Applied Quantum Algorithms \texorpdfstring{$\langle \mathrm{aQa}^\mathrm{L}\rangle$}{<aQa^L>}, LIACS \& LION, Leiden University, The Netherlands}

\author{James Enouen}
\affiliation{Department of Computer Science, University of Southern California, Los Angeles, CA 90089, USA}

\author{Guillaume~Bornet}
\affiliation{Princeton University, Department of Electrical and Computer Engineering, Princeton, New Jersey 08544, USA}

\author{Thierry~Lahaye}
\affiliation{Université Paris-Saclay, Institut d’Optique Graduate School,
CNRS, Laboratoire Charles Fabry, 91127 Palaiseau Cedex, France}

\author{Antoine~Browaeys}
\affiliation{Université Paris-Saclay, Institut d’Optique Graduate School,
CNRS, Laboratoire Charles Fabry, 91127 Palaiseau Cedex, France}

\author{Antoine~Georges}
\affiliation{Collège de France, PSL University, 11 place Marcelin Berthelot, 75005 Paris, France}
\affiliation{Center for Computational Quantum Physics, Flatiron Institute, 162 Fifth Avenue, New York, NY 10010, USA}
\affiliation{CPHT, CNRS, École Polytechnique, IP Paris, F-91128 Palaiseau, France}
\affiliation{DQMP, Université de Genève, 24 quai Ernest Ansermet, CH-1211 Genève, Switzerland}

\author{Anna~Dawid}
\email{a.m.dawid@liacs.leidenuniv.nl}
\affiliation{Applied Quantum Algorithms \texorpdfstring{$\langle \mathrm{aQa}^\mathrm{L}\rangle$}{<aQa^L>}, LIACS \& LION, Leiden University, The Netherlands}

\date{\today}

\begin{abstract}
Detecting phases of matter in general relies on identifying the correct order parameter - a task that remains notoriously difficult for unknown transitions and traditionally is guided by physical intuition and educated guess. Neural networks have recently offered an alternative route by locating phase transitions in known models without any a priori physical knowledge. Yet these approaches remain black boxes and only identify phases without elucidating their properties. Moreover, they often struggle when confronted with realistic, noisy experimental data, which constitute the ultimate testbed for automated methods in physics.
Here, we bridge these perspectives by introducing TetrisCNN, a convolutional architecture with parallel branches of differently shaped filters, reminiscent of Tetris blocks, that learns sparse, interpretable latent representations directly in terms of spin correlators. Applied to experimental snapshots of two-dimensional Ising and XY quantum simulators measured in multiple bases, the network not only detects phase transitions and crossovers but also expresses its latent representation and decision boundaries as symbolic formulas built from experimentally measurable spin correlators. This framework opens the way to integrating interpretable neural networks with quantum simulators to uncover and understand new phases of matter.

\end{abstract}

\maketitle

\section{Introduction}

Understanding complex quantum systems with long-range interactions and strong entanglement remains a central challenge of modern quantum physics, particularly as classical simulation becomes computationally prohibitive. Programmable quantum simulators \cite{Gross2017, Kjaergaard20AnnuRev, Browaeys20NatPhys, Monroe21RevModPhys} provide a remarkable experimental alternative, granting access to regimes beyond the reach of conventional numerical methods and potentially uncovering new phases of matter thanks to their tunability \cite{barredo2018synthetic-665}. However, identifying suitable order parameters for previously unknown phase transitions remains nontrivial and often depends on expert intuition. The task requires navigating an exponentially large Hilbert space and analyzing the symmetries of the system, guided by physical insight and educated guesses.

In parallel, machine learning has an increasing impact on quantum sciences \cite{carrasquilla_machine_2020, Krenn23MLforQT, medvidovic2024NQS, dawid2025machine-b47, schlmer2025machine-7d6}, mirroring its influence on industry~\cite{dawid2024introduction-db8}. In particular, neural networks have emerged as powerful tools for detecting phases of matter \cite{Carrasquilla17NatPhys, van_nieuwenburg_learning_2017, wetzel_unsupervised_2017, greplova_unsupervised_2020, bohrdt2021analyzing, patel2022unsupervised, arnold2024generative, kim2025attention, malyshev2026distance-fb9}, and they are especially promising in unsupervised settings, where no prior knowledge of the system or its order parameters is assumed. In principle, such techniques could enable the direct identification of novel and exotic phases from experimental data, including ones that may have escaped human intuition. 
In practice, however, machine learning has uncovered few unexpected phases, with notable examples using autoencoders \cite{kottmann_unsupervised_2020} and support vector machines \cite{Liu2021}. Such discoveries remain rare partly because machine learning is seldom applied directly to experimental quantum-simulator data \cite{schlmer2025machine-7d6}.

One reason for the challenging nature of learning from experiments is that, in practice, experimental constraints often restrict the set of accessible observables \cite{barthel2018fundamental-75d}, preventing access to full information about the quantum state. Moreover, real-world measurements introduce noise and finite-size effects that can obscure phase transitions, and put limits on system sizes \cite{ziv2025unsupervised-bb2}. These challenges complicate both training and the interpretation of results, and have made supervised learning the most common approach when analyzing experimental data \cite{Rem19, Khatami20, Kaming2021, link2023machine, miles2023machine}. However, supervised models rely on ground truth and struggle with out-of-distribution generalization, limiting their potential for scientific discovery. Additional concerns include the stability of learning models against noise in experimental settings that can act like adversarial perturbations \cite{zhang2022experimental-ad8} and therefore fatally derail network predictions \cite{madry2018towards}.

\begin{figure*}[t]
    \centering
    \includegraphics[width=0.99\textwidth]{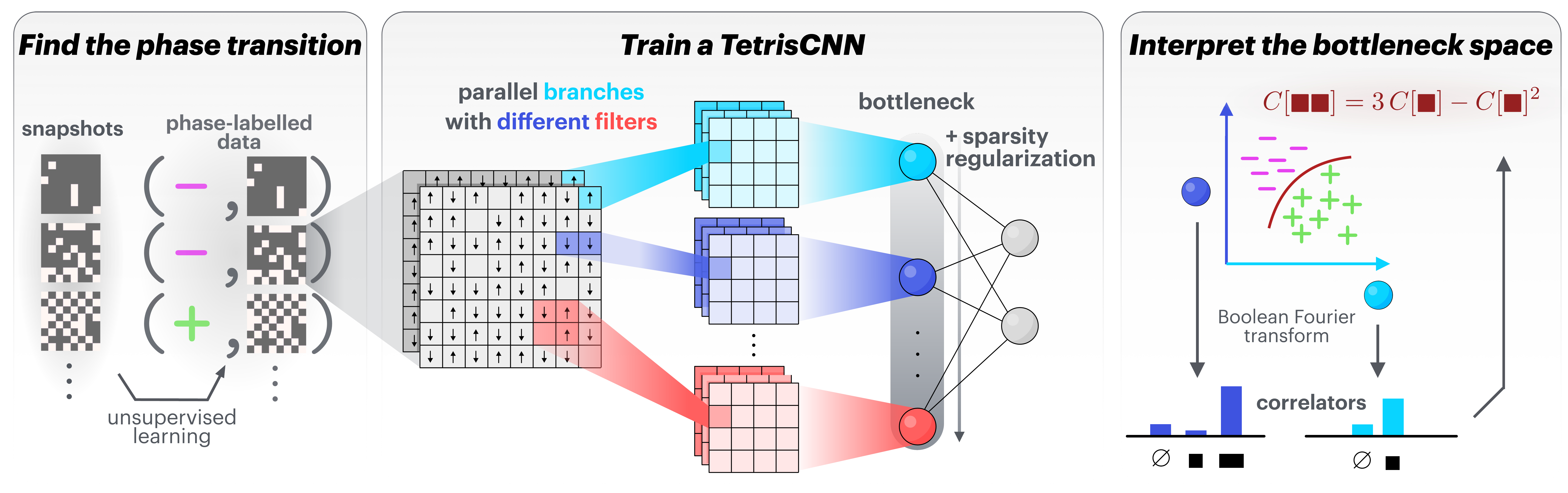}
    \caption{\figcap{TetrisCNN processes snapshots through parallel branches with differently shaped filters that map to spin correlators.} Sparsity suppresses redundant or task-irrelevant branches, yielding a low-dimensional interpretable bottleneck from which the decision boundary can be expressed symbolically in terms of spin correlators.
    }
    \label{fig:intro}
\end{figure*}

Even when unsupervised detection schemes are successfully applied directly to experimental data \cite{Kaming2021, yu2022experimental-97c, miles2023machine, ziv2025unsupervised-bb2, schoulepnikoff2025interpretable-3a6, schoulepnikoff2026discovering-e9b, sadoune2026learning-c10}, the resulting machine learning models typically provide little physical insight, with only a few notable exceptions~\cite{miles2023machine, sadoune2026learning-c10,schoulepnikoff2025interpretable-3a6, schoulepnikoff2026discovering-e9b}. 
Accurate prediction alone does not encourage neural networks to find interpretable descriptions: they lack an intrinsic drive to “compress knowledge” \cite{elhage2022toy-ef0, klindt2025from-ada, vafa2024evaluating-9f8, pmlr-v267-vafa25a}. 
Extracting reliable physical insight is further complicated in experimental settings, where explainability methods can be even more sensitive to noise than the underlying predictions~\cite{ghorbani_interpretation_2019, cybinski2025ssh}, while the Rashomon effect (the coexistence of multiple, equally predictive yet qualitatively distinct models, whose number grows with noise) undermines the stability of inferred conclusions~\cite{pmlr-v235-rudin24a}.
An ultimate goal of machine learning for phases of matter is therefore to make such automated approaches interpretable \cite{Dawid20NJP, wetzel_discovering_2020, Dawid2021Hessian, arnold_interpretable_2021, arnold2022replacing, arnold2023fisher, wetzel2024closedform, zhan2024learning, cybinski2025ssh, ju2025interpreting-8a1}, especially for experimental data, enabling not only the detection of phases of matter but also an understanding of their physical properties, in particular the underlying order parameters~\cite{miles2023machine, sadoune2026learning-c10, wetzel2017orderparams, Greitemann19probing, Liu19interpretable, Greitemann19identification, sadoune2023unsupinterpet, sadoune2025human-machine-a0b, cole_quantitative_2021, Miles2021CCNN, striegel2023machine-de1, schlomer2023fluctuationinterpret, cao2024unveiling, suresh2024ctransformer}.

To address these challenges, we introduce TetrisCNN, an interpretable convolutional neural-network architecture for identifying the spin correlators underlying phase transitions and crossovers directly from experimental snapshots. The key idea is to provide the network with an explicit drive to compress knowledge. TetrisCNN uses parallel convolutional branches that represent candidate local correlators at different spatial scales and geometries, with the resulting collection of differently shaped filters motivating the name ``TetrisCNN''. Sparse regularization then automatically selects a small subset of correlators sufficient for a given learning task. This design enables the latent representation and the resulting decision boundaries to be expressed analytically in terms of experimentally measurable spin correlators, providing a missing ingredient of neural network-based approaches: interpretability. We apply TetrisCNN to experimental two-dimensional (2D) Rydberg-array data that realize the Ising \cite{Scholl2021qsimIsing} and XY \cite{Chen2023qsimXY} models, where it recovers sparse, physically meaningful representations of the observed transitions and reveals the correlators underlying the network predictions. Throughout this work, we use \emph{transition} broadly for pronounced changes in the system's behavior, encompassing both phase transitions and crossovers. Beyond accurately detecting transitions, TetrisCNN reveals why the predictions are made, allowing machine-learning conclusions to be analyzed and challenged using theoretical arguments.

The manuscript is organized as follows. We first introduce the experimental Ising and XY datasets, the learning task, and the TetrisCNN architecture, including the sparse bottleneck and Boolean-Fourier mapping to spin correlators (\seclab~\ref{ss:Hamiltonians}--\ref{ss:interpretable_Tetris}). We then present the results in \seclab~\ref{sec:results}, discuss their physical interpretation in \seclab~\ref{sec:discussion}, and compare TetrisCNN with related interpretable approaches in \seclab~\ref{sec:related_work}. We conclude in \seclab~\ref{sec:conclusion}. Implementation details and all architectural and numerical settings are provided in \apps~\ref{app:data_preparation}--\ref{app:SR}, together with the open-source code and experimental datasets at GitHub~\cite{our_github_2026}.

\begin{figure*}[t]
    \centering
\includegraphics[width=\textwidth]{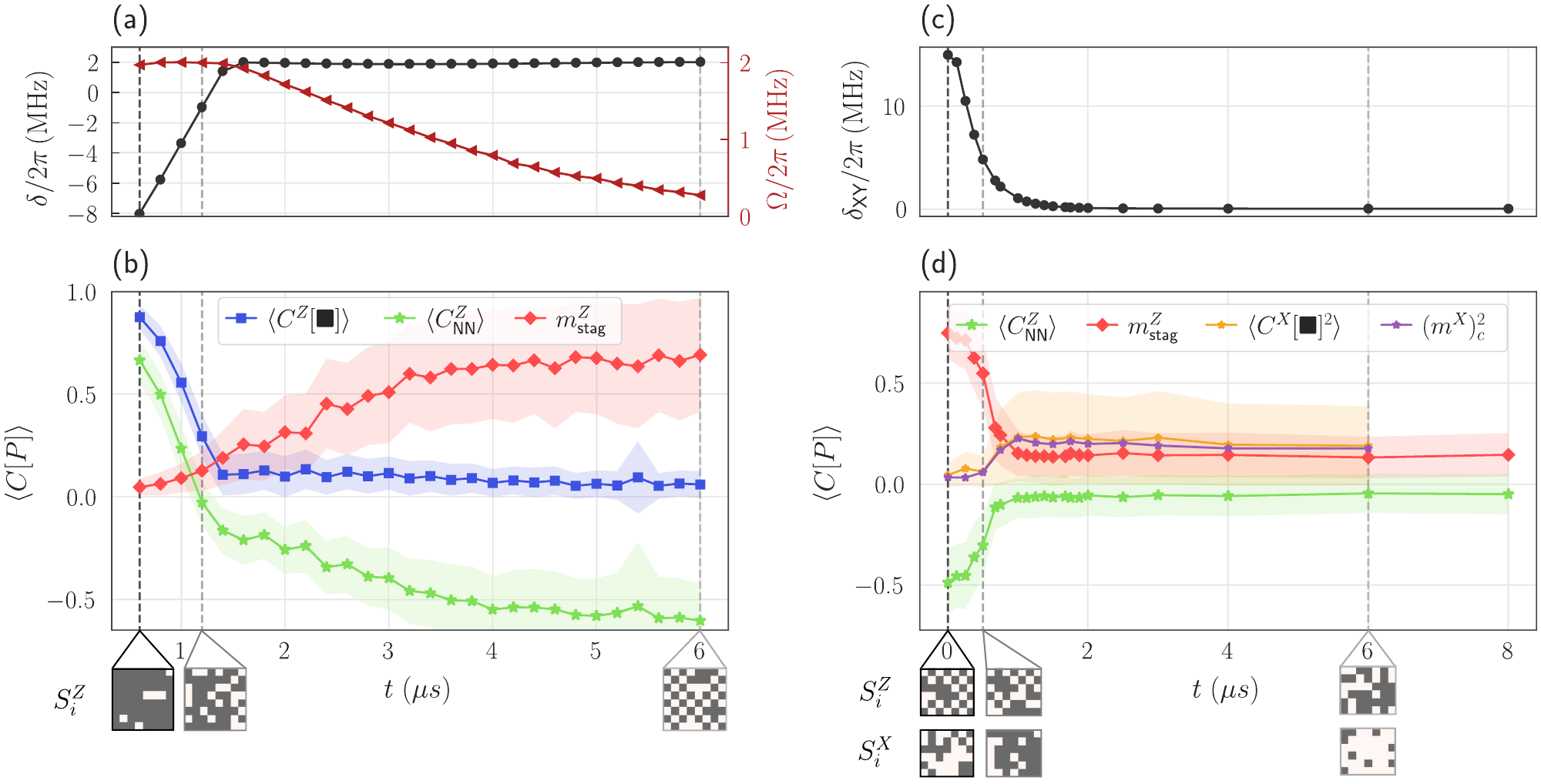}
    \caption{\figcap{Experimental quantum data.} (a),(c) Experimental protocols for the Ising and XY datasets, including control parameters, measurement points, and representative snapshots from the $8\times8$ Ising~\cite{Scholl2021qsimIsing} and $6\times7$ XY~\cite{Chen2023qsimXY} Rydberg-atom quantum simulators. 
    (b),(d) Selected snapshot-averaged spin correlators, cf. \eq~\eqref{eq:C(P)}, with shaded regions indicating their standard deviation across the sweeps: 
    $m_\text{stag}=C_A[\bs]-C_B[\bs]$, with $A,B$ being the two sublattices,
    $C_\text{NN} = C_{\mathrm{rot}} [\bs \bs] = \frac{1}{2}(C[\bs \bs]+C[\sus{\bs \\ \bs}])$, 
    $(m^X)^2$ is in-plane ferromagnetic magnetization squared, cf.~\eq~\eqref{eq:mX2}.
    }
    \label{fig:quantum_data}
\end{figure*}

\section{Methods}\label{sec:methods}

\subsection{Quantum data}\label{ss:Hamiltonians}

We showcase the usefulness of TetrisCNN on the example of interpretable detection of phases of matter from raw experimental projective measurements (snapshots) taken on the Rydberg quantum simulator of two quantum spin Hamiltonians: 2D Ising and XY models.

\paragraph{Rydberg quantum simulators} Among the most promising quantum simulation platforms are Rydberg-atom arrays~\cite{Labuhn2016theoryexp, Browaeys20NatPhys}. Atoms trapped in optical tweezers form highly controllable pseudospin-$\frac12$ qubits that can be individually positioned, addressed, and measured with single-site resolution. Quantum correlations and entanglement are generated via strong electric dipole--dipole interactions, giving rise to the Rydberg blockade mechanism that suppresses simultaneous excitation of nearby atoms. Crucially, the geometry, interaction range, and driving protocols are highly tunable and programmable, which (after accounting for experimental imperfections \cite{simard2025learning-205}) makes them exciting platforms for exploring emergent many-body phenomena and realizing and studying Hamiltonians that are inaccessible to classical numerical methods.

\paragraph{2D Ising Hamiltonian} 
A 2D array of Rydberg atoms coupled via repulsive van der Waals interactions naturally realizes the Hamiltonian of the transverse-field Ising model (TFIM) when driven to Rydberg states~\cite{Scholl2021qsimIsing}:
\begin{equation}\label{eq:TFIM_ham}
    \hat{H}_{\rm Ising} = \sum_{i<j} U_{ij} n_i n_j + \frac{\hbar \transverse}{2} \sum_i \sigma_i^X - \hbar \longitudinal \sum_i n_i \,,
\end{equation}
where the Rydberg and ground states are the (pseudo)spin states $\spinup = |75S_{1/2}, m_J=1/2\rangle$ and $\spindown = |5S_{1/2}, F=2, m_F=2\rangle$, respectively, with $F$ and $m_F$ denoting the hyperfine quantum numbers and $m_J$ the magnetic quantum number of the fine structure. The interaction strength between atoms is given by the van der Waals potential $U_{ij}=C_6/r_{ij}^6$, where $C_6$ is the van der Waals coefficient and $r_{ij}$ is the distance between atoms $i$ and $j$. Here $n_i=\lvert\uparrow\rangle\!\langle\uparrow\rvert_i=(1+\sigma_i^Z)/2$, $\sigma_i$ are the Pauli matrices, and $\hbar$ is the reduced Planck constant. The two spin states are coherently coupled by a laser field with Rabi frequency $\transverse$ and detuning $\longitudinal$, which play the roles of the transverse and longitudinal fields, respectively. For a square lattice with lattice spacing $a=10\,\mu\mathrm{m}$ and atoms excited to the $n=75$ Rydberg state, the nearest-neighbor interaction strength is $U/h \approx 1.95\,\mathrm{MHz}$.

For this Hamiltonian, snapshots are measured in the $Z$ basis for $t\in[0.6,6]\,\mu\mathrm{s}$ at intervals of $0.2\,\mu\mathrm{s}$. During this protocol, $\delta$ and $\Omega$ are varied quasi-adiabatically, as shown in \fig~\ref{fig:quantum_data}(a) (see \reflab~\cite{Scholl2021qsimIsing} for details). The system is initially prepared in the $Z$-polarized paramagnetic ground state $|\downarrow\downarrow\ldots\downarrow\rangle$. As the detuning is increased while the transverse field remains finite, the system first undergoes a smooth crossover out of this nearly fully polarized state
and enters a more strongly mixed paramagnetic regime. At later times, the transverse field is reduced and the protocol crosses the paramagnetic-to-antiferromagnetic quantum phase transition, leading to
the growth of staggered magnetization.

\paragraph{XY Hamiltonian} 
Another experimental setup \cite{Chen2023qsimXY}, whose measurements we study, consists of a 2D rectangular lattice of ${}^{87}$Rb atoms in an optical $6 \times 7$ tweezer array, where an effective spin-1/2 is encoded in the Rydberg states $\spinup = |60S_{1/2}\rangle$ and $\spindown = |60P_{1/2}\rangle$. Resonant dipole--dipole interactions allow the realization of a long-range dipolar XY Hamiltonian supplemented by a staggered longitudinal field, leading to the total Hamiltonian
\begin{equation}
\label{eq:XY_ham}
\begin{aligned}
\hat{H}_{\mathrm{XY},\deltaxy}
={}& -\frac{J}{2} \sum_{i<j} \frac{a^3}{r_{ij}^3}
\left( \sigma_i^X \sigma_j^X + \sigma_i^Y \sigma_j^Y \right) \\
& + \hbar \deltaxy
\sum_i \frac{1-(-1)^{i_x+i_y}}{2}\,n_i \,.
\end{aligned}
\end{equation}
Here, $J/h=0.77\,\mathrm{MHz}$, the lattice spacing is $a=12.5\,\mu\mathrm{m}$, and a magnetic field perpendicular to the lattice plane defines the quantization axis, resulting in isotropic dipolar interactions. The integers $(i_x,i_y)$ denote the coordinates of site $i$ on the square lattice, such that $(-1)^{i_x+i_y}$ alternates between the two sublattices. With the convention $(-1)^{i_x+i_y}=+1$ on sublattice $A$ and $-1$ on sublattice $B$, the staggered-field term vanishes on $A$ and equals
$\hbar \deltaxy n_i$ on $B$, where $n_i=(1+\sigma_i^Z)/2$. It therefore describes the staggered light shift used experimentally to prepare the initial N\'eel state.

For a sufficiently large positive value of $\deltaxy$, the initial N\'eel state approximates the ground state of $\hat H_{\mathrm{XY},\deltaxy}$. Reducing the staggered field toward zero following an approximately adiabatic exponential ramp, as shown in \fig~\ref{fig:quantum_data}(c), connects this state to a low-energy state of
the XY Hamiltonian with ferromagnetic correlations in the $XY$ plane. Conversely, for a sufficiently large negative staggered field, the same N\'eel configuration approximates the highest-energy state, and the
corresponding ramp produces antiferromagnetic $XY$ correlations \cite{Chen2023qsimXY}. In this work, we analyze the ferromagnetic protocol. Snapshots are measured in both the $X$ and $Z$ bases for $t\in[0,8]\,\mu\mathrm{s}$ at irregular intervals.

\paragraph{Working with experimental data vs simulated data} 
Most deep learning studies in quantum physics rely on numerically simulated ground states.
TetrisCNN was also initially developed as a proof of concept on numerically simulated datasets, including the one-dimensional (1D) TFIM and the 2D Ising lattice gauge theory \cite{cybinski2024tetriscnn}.
We believe, however, that applying machine learning methods to experimental or noisy data is the ultimate test for their utility and robustness \cite{cybinski2025ssh}.
Therefore, here we analyze raw snapshots from quantum simulation experiments. These configurations are produced by finite-time approximately adiabatic sweeps and therefore represent non-equilibrium dynamical states rather than exact ground states. The data additionally contain realistic experimental imperfections, including state preparation, detection, and basis-rotation errors, inhomogeneous fields, interaction disorder from atomic-position fluctuations, and decoherence during the ramps. While local errors lead to bounded errors in local observables, these imperfections reduce the fidelity of the quantum many-body state and pose increasing challenges for preparing and characterizing larger systems. Since the experiments are performed on finite lattices with open boundaries, edge and finite-size effects also influence the observed ordering patterns. Finally, comparisons with classical equilibrium ensembles \cite{Scholl2021qsimIsing} reveal systematic biases toward more ordered configurations, indicating that the data reflect specific noisy quantum dynamics rather than thermal or idealized equilibrium samples. 

\paragraph{Datasets}
We now take a machine learning perspective. Projective measurement snapshots from the models above are taken as the network input $x^{(n)}=\{\snapshot_i^{b}\}_{i=1}^{\systemsize}$, with $i$ indexing the spatial grid and $b\in\{X,Z\}$ denoting the measurement basis. When both bases are available, we randomly combine snapshots acquired at the same sweep point into two channels as $x^{(n)}=\{(\snapshot_i^{X},\snapshot_i^{Z})\}_{i=1}^{\systemsize}$. Each snapshot has a corresponding label $y^{(n)}$ that depends on the learning task, which we specify further in \seclab~\ref{ss:task}.
Snapshots and labels are collected in a dataset $\mathcal \dataset \equiv \{(x^{(n)},y^{(n)})\}_{n=1}^{\datasize}$. We focus on 2D square grids, so the data is represented as a tensor of shape $[N,C,D_1,D_2]$ with $D=D_1\times D_2$ the number of spatial sites, $C=1$ by default, and $C=2$ if we consider two bases simultaneously.

$\mathcal \dataset$ is divided into training and validation sets in a 7:3 ratio. We measure evaluation metrics on the validation set for early stopping, and to compare the fits of different models. We do not use a test set because we are not interested in benchmarking; rather, we use machine learning as a supervised subroutine in a broader unsupervised discovery process (i.e., evaluation metrics on the validation set should not be interpreted as expected model performance, as is common in machine learning). More information on dataset preparation is in \app~\ref{app:data_preparation}.

\paragraph{Order parameters}
In general, quantum systems exhibit a wide variety of types of \textit{order}. The Landau paradigm defines order by symmetry breaking \cite{Landau1937, Hohenberg15} (e.g. in the TFI and XY model above); BKT order relates to binding/unbinding of topological defects \cite{Berezinskii1971, KosterlitzThouless1973} (e.g. the AFM variant of the XY model); topological order is characterized by long-range entanglement and topology-dependent ground state degeneracy \cite{Wen1990, WenNiu1990} (e.g. the toric code \cite{Kitaev2003}). Order can be even more unconventional~\cite{nijs1989preroughening-9e3,hikihara2008vector-f3f} and novel types of order may yet be discovered. We desire a machine learning approach that provides insight regardless of the system's order -- here, we take a step towards this goal.

\paragraph{Spin correlators}
For a machine learning algorithm to provide insight, it needs to be expressed in a language we can recognize. A general way to characterize order in many-body systems is through spin correlators or $n$-point functions. For example, ferromagnetic order can be characterized by the one-point function, or magnetization, $\overline{S_i} = m$, where the overline bar denotes a spatial average over indices $i$ --
or using long-range behavior of the two-point function $\lim_{|i-j|\rightarrow\infty}\overline{S_{i} S_{j}}$. Staggered magnetization can be written as $\overline{S_A}-\overline{S_B}$ where $A$ and $B$ are two sublattices. 
In general, correlators are characterized by the number of spins they involve. More complicated phases, however, involve correlators in \textit{specific patterns} (e.g. staggered, rhombic or striated patterns \cite{kalinowski2022bulk-efc, orourke2023entanglement-8a5}). 

We therefore represent a spin correlator by a corresponding geometrical pattern $P$ that specifies the relative position of spins whose product is taken. For a given snapshot, let $\mathcal{T}_P$ denote all translations of $P$ that fit within the physical system. We define the corresponding single-snapshot spin correlator as
\begin{equation}
\boxed{
        C[P](S)=
    \frac{1}{|\mathcal{T}_P|}\sum_T \prod_{i\in T(P)} S_{i}
}
    \label{eq:C(P)}
\end{equation}
that is, the product of the spins within a pattern $P$ translated by $T$. Throughout, sums over $T$ are always defined over all possible translations of the pattern within the snapshot, $T\in\mathcal{T}_P$.
From here onwards, we shall omit the dependence on spins $(S)$ to avoid clutter.
For example, 
$P=\bs\bs=\{ 0,1\}$
defines the horizontal nearest-neighbor spin pair, and therefore $C[\bs\bs] =\frac{1}{D-1}\sum_i S_{i} S_{i+1}= \overline{S_{i} S_{i+1}}$, for a snapshot of a one-dimensional system of size $D$. Similarly, the single-site pattern $P=\bs$ gives the snapshot magnetization, $C[\bs]=m$.

When no lattice orientation is physically distinguished, it is useful to consider correlators that are invariant under lattice rotations. We define the rotationally invariant correlator associated with $P$ by averaging over its distinct rotations on a square lattice,
\begin{equation}
\boxed{
C_{\mathrm{rot}}[P] = \frac{1}{|\mathcal{R}_P|} \sum_{R} C[R],
}
\label{eq:C_rot(P)}
\end{equation}
where
$R\in \mathcal{R}_P$ 
are all distinct $90^\circ$ rotations of $P$, corresponding to the $C_4$ symmetry group.
For example,
$C_{\mathrm{rot}}[\bs\bs]
=
\frac{1}{2}
\left(
C[\bs\bs]
+
C\!\left[\sus{\bs\\\bs}\right]
\right)$.

Throughout this work, $C[P]$ denotes a quantity evaluated on a single
snapshot. We denote its expectation value over experimental snapshots by
$\langle C[P]\rangle$ (and analogously $m$ and $\langle m\rangle$ for the
magnetization). Several such correlators are plotted in
\fig~\ref{fig:quantum_data}(b) and (d) for the TFIM and XY datasets.

\subsection{Task}\label{ss:task}

Several unsupervised methods for finding phase transitions have been proposed in the literature. These include clustering techniques applied within a low-dimensional space, which can be obtained by dimensionality-reduction techniques \cite{Wang16}, diffusion maps \cite{Lidiak20PRL}, or autoencoders \cite{wetzel_unsupervised_2017, kottmann_unsupervised_2020, kottmann2021variational-d21, schoulepnikoff2025interpretable-3a6, moller2026learning-648, schoulepnikoff2026discovering-e9b}. Other approaches include learning by confusion \cite{van_nieuwenburg_learning_2017}, prediction-divergence method \cite{schafer_vector_2019, greplova_unsupervised_2020, arnold_interpretable_2021}, generative modeling \cite{arnold2024generative}, and distance learning \cite{ziv2025unsupervised-bb2, malyshev2026distance-fb9}. 

TetrisCNN is an architecture that can serve as a drop-in replacement in a large number of methods: autoencoders, prediction divergence (\app~\ref{app:task_dependence}), learning by confusion, distance learning, and supervised classification. For clarity and simplicity, we present in this work how TetrisCNN can be used for supervised classification. The estimated phase boundary is determined separately in a prior stage: we use and compare several methods and adopt the transition point that is most frequently observed among these predictions.

When using TetrisCNN for supervised learning, we employ the cross-entropy loss:
$$\mathcal{L}_{\text{CEL}} = -\sum_{n,i} y_i^{(n)}\log \hat y_i^{(n)},
$$ 
with $i\in\{0,1\}$ the class label and $\hat {y} $ a softmax of the final layer. Task performance is measured by accuracy, the fraction of snapshots correctly classified. 

Note that the network here makes predictions on single snapshots, which sometimes limits the maximum accuracy it can achieve. For example, if the same spin configuration appears in different phases, the network cannot make correct predictions for all of them. Moreover, networks trained on single snapshots cannot compute directly so-called ensemble connected correlators \cite{Chalopin26connectedcorrs} such as $\langle S_i S_{i+1} \rangle_\mathrm{c} = \langle S_i S_{i+1} \rangle - \langle S_i \rangle \langle S_{i+1} \rangle$, as in the in-plane magnetization squared $(m^X)_c^2$ in \fig~\ref{fig:quantum_data}(d), cf. \eq~\eqref{eq:mX2}.

\subsection{Interpretable architecture of TetrisCNN}\label{ss:interpretable_Tetris}

\paragraph{Interpretability and expressivity}
Neural networks (NNs) stand out among machine learning methods by being both trainable and highly expressive. This \textit{expressivity} is associated with network depth, width, and total parameter count \cite{eldan2015power-0e3, kaplan2020scaling-f4c}. Neural networks also excel at representation, or feature learning \cite{lecun_deep_2015,pmlr-v267-wilson25a}. Extracting or \textit{interpreting} these features from the network, however, is typically extremely difficult.
Hence, it is commonly assumed that expressivity and {interpretability} are inherently opposed \cite{NEURIPS2021_251bd044, enouen2022sparse, rudin2022interpretable-939}. Our network design, TetrisCNN, showcases how this is not necessarily true for Boolean-valued data (e.g., spins). 

This is achieved using a combination of elements that conspire to enable interpretation -- here defined in a broad sense. We briefly explain each design element below, including its contribution to interpretability. We emphasize the generality of our approach; TetrisCNN provides a drop-in replacement for any neural network that was heretofore a black box.

\begin{figure}
    \centering
    \includegraphics[width=0.98\columnwidth]{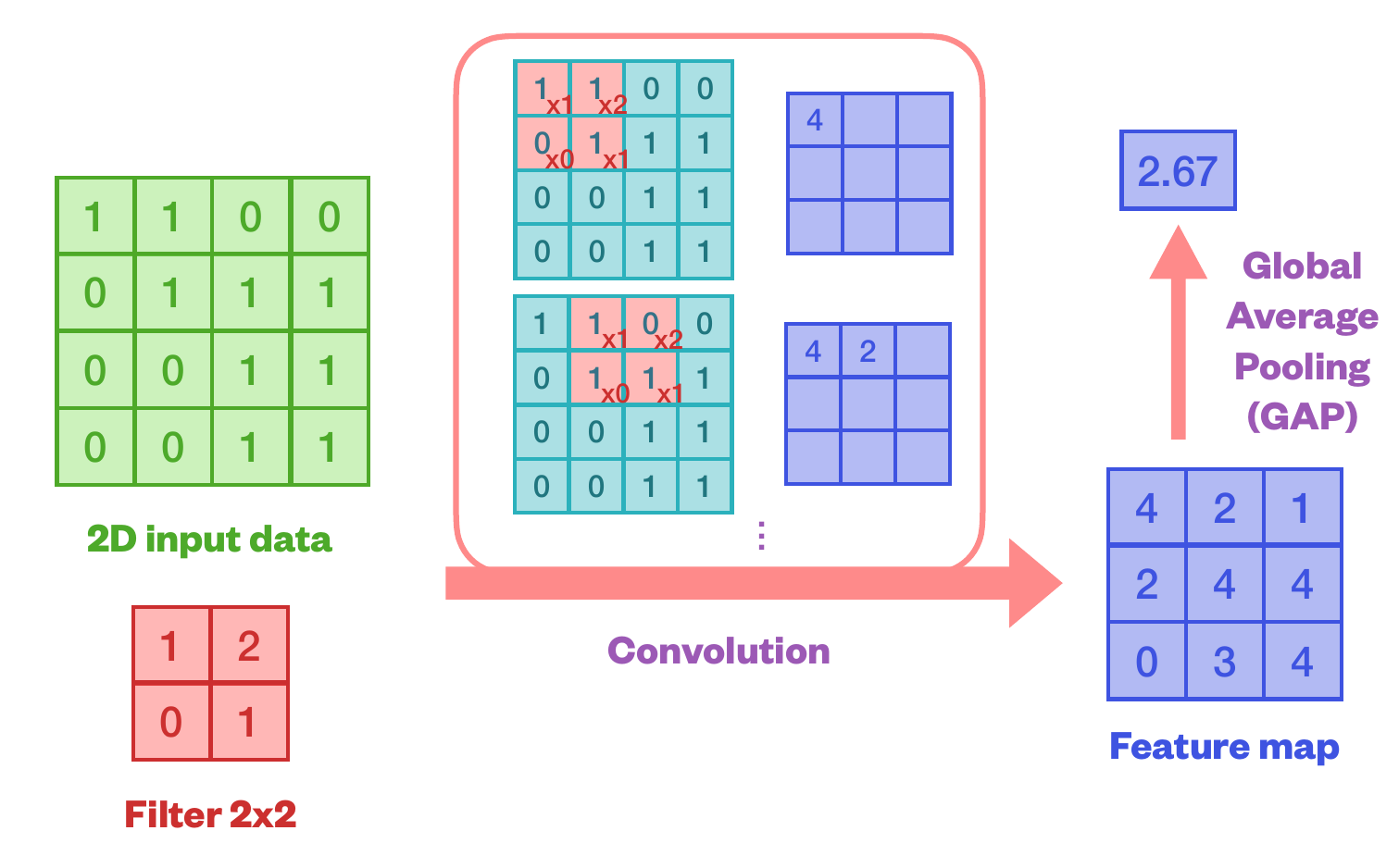}
    \caption{\figcap{Example of convolution with a $2 \times 2$ filter, followed by global average pooling (GAP).} 
    The filter is applied across the input to produce a feature map, which GAP reduces to a scalar.
    }
    \label{fig:CNN}
\end{figure}

\paragraph{Locality and spatial symmetries}
Our data is defined on a 2D square lattice, implying an approximate translational symmetry (an infinite lattice would be fully symmetric). The physics defined by \eqs~\eqref{eq:TFIM_ham} and \eqref{eq:XY_ham} is also \textit{mostly} local, interactions decaying as $1/r_{ij}^{6}$ and $1/r_{ij}^{3}$, respectively. This motivates the use of \textit{convolutional layers}, which enable highly efficient learning on data with these properties \cite{LeCun98}.
Reusing the notation in \eq~\eqref{eq:C(P)}, we express a convolution in a non-standard form using a pattern $P$:
\begin{equation}
\operatorname{conv}[P]_T = \sum_{j\in P} W_{j}  S_{T(j)}\,.
\label{eq:conv}
\end{equation}
Here, $S$ is the full input spin configuration (snapshot), 
while pattern $P$ specifies the support of the convolution. We again omit the dependence on $S$ on the left-hand side to avoid clutter.
$W_j$ is its associated trainable weight, 
and $T(j)$ is the corresponding translated lattice site.
The pattern $P$, together with the weights $\{W_j\}_{j \in P}$, defines a \textit{filter}: a local function that is applied with the same weights at every allowed translation $T$, as illustrated in \fig~\ref{fig:CNN} for a $2 \times 2$ filter. Thus, \eq~\eqref{eq:conv} is the filter output at translation $T$ and depends only on the spins, and $\{S_i\}_{i\in T(P)}$.
Applying the same filter at every allowed translation of $P$ within the snapshot produces a \textit{feature map}.

Convolutions are typically paired with pooling layers, which aggregate or downsample the resulting feature map. Of particular interest here is global average pooling (GAP), which averages the feature map over all spatial positions (see the final arrow in \fig~\ref{fig:CNN}).
The convolution in \eq~\eqref{eq:conv} is
\textit{equivariant}: translating the input translates the feature map. GAP then removes the spatial index, making the pooled output translation \textit{invariant} (up to finite-size boundary effects). 

While standard convolutions (\eq~\eqref{eq:conv}) capture translational symmetry, the $C_4$ rotational symmetry of the square lattice can be accounted for with group-convolutions. Following \cite{cohen2016group}, we implement this by applying each learned filter in all distinct rotations 
$R\in\mathcal R_P$
, rotating its pattern and weights together, and averaging the resulting feature maps:
\begin{equation}
\operatorname{conv}_{\mathrm{rot}}[P]_T \equiv \frac{1}{|\mathcal R_P|} \sum_{R} \operatorname{conv}[R]_T.
\label{eq:rot_conv}
\end{equation}
Here, $R\in\mathcal R_P$ denotes all distinct $90^\circ$ rotations of $P$, and 
$\operatorname{conv}[R]$
denotes the correspondingly rotated version of the same filter, including its weights. This makes the convolutional feature map rotationally equivariant, and its spatially pooled output rotationally invariant.

\paragraph{Interpreting convolutions with correlators}

Previous work \cite{wetzel2017orderparams} showed 
that when any \textit{function of a convolution} (e.g. a convolution followed by a nonlinearity) is combined with global average pooling (GAP), a remarkable interpretation can be made for \textit{Boolean-valued} spin configurations 
$S\in\{\pm1\}^D$
Let 
$f[P]_T$ 
denote the output at the translated pattern 
$T$
of such a local function with receptive field $P$.
Using a Boolean Fourier expansion \cite{odonnell2014analysis-ead, schurov2025learning-62d, dschl2025importance-82b, pmlr-v284-nicolau25a} (cf. \app~\ref{app:boolean_fourier}), one can show that
\begin{equation}
\begin{aligned}
\frac{1}{|\mathcal{T}_P|} \sum_{T} f[P]_T &= \sum_{P'\subseteq P} c_{P'}(f)\,C[P']\,,
\end{aligned}
\label{eq:sum_of_correlators}
\end{equation}
where 
$c_{P'}(f)\in\mathbb{R}$
are coefficients determined by the function 
$f[P]$, and $P'\subseteq P$ runs over all subpatterns of $P$. Thus, any local function of Boolean-valued spins, when combined with GAP, produces \textit{a linear combination of the spin correlators} $C[P']$ defined in \eq~\eqref{eq:C(P)}.
This was previously shown in \reflab~\cite{wetzel2017orderparams} for small patterns ($\bs$ and $\bs\bs$), e.g., for $P = \bs\bs$, the GAP output takes the form 
$c_{\emptyset} + c_1 C[\bs] + c_2 C[\bs\bs]$
(up to open boundary effects).
\Eq~\eqref{eq:sum_of_correlators} generalizes this result to arbitrary patterns (correcting a minor theoretical error in their derivation in the process, replacing it by a principled foundation using Boolean Fourier analysis). 

It should be noted that $\{c_{P'}\}$ are not directly available, but can be obtained post hoc by simply taking a linear regression fit on the output using $\{C[P']\}$. 
This decomposes the learned function into physically meaningful spin correlators. 
We also note that this procedure is efficient up to a $3 \times 3$ filter but becomes expensive for larger filters, due to exponential scaling of the number of subpatterns $P'\subseteq P$ as $|P|$ grows (400 and 57,856 translationally symmetric subpatterns for $3 \times 3$ and $4 \times 4$ filters, respectively; see \app~\ref{app:counting_correlators}).

\paragraph{Interpretable branch definition} With this result, we now define a TetrisCNN \textit{branch} as
\begin{equation}\label{eq:branch}
\begin{aligned}
z[P]&\equiv \frac{1}{|\mathcal{T}_P|} \sum_{T} \Big[ \operatorname{conv} [\bs] \circ\phi\circ\operatorname{conv} [P]  \Big]_T.
\end{aligned}
\end{equation}
This sequence of operations is designed to maximize expressivity through layers and channels, while preserving the expansion in \eq~\eqref{eq:sum_of_correlators}. Specifically, the first convolution $\operatorname{conv}[P]$ sets the branch's receptive field to $P$, and produces 32 output channels.
The subsequent $1 \times 1$ convolution, $\operatorname{conv}[\bs]$, combines these channels locally without enlarging the receptive field. The function $\phi$ is an elementwise nonlinearity, chosen here to be ReLU. Finally, GAP averages the resulting feature map over all translations $T$, producing the scalar branch output $z[P]$.
As examples, we show several $z[P]$ in \tab~\ref{tab:Tetris_correlators} for patterns of increasing size, along with the correlators that each $z[P]$ can compute. 

\begin{table}[ht]
    \centering
\resizebox{\columnwidth}{!}{%
\begin{tabular}{l|l|l}
\multicolumn{1}{l}{\textbf{}} 
& \multicolumn{1}{l}{\textbf{Convolution filter} $\operatorname{conv}[P]$} 
& \multicolumn{1}{l}{\textbf{Correlators} $C[P'], P'\subseteq P$} \\
\hline 

$z_1$ 
& $\text{conv}[{\blacksquare}]_i = {W_0}\textcolor{MyBlue}{S_i}$ 
& 
$
\begin{array}{lcl}
C[\emptyset] &=& \text{const.} \\
C[\textcolor{MyBlue}{\blacksquare}] 
&=& \overline{\textcolor{MyBlue}{S_i}}
\end{array}
$
\\

\hline 

$z_2$ 
& $\text{conv}[{\blacksquare}{\blacksquare}]_i =
{W_0}\textcolor{MyBlue}{S_i}
+ {W_1}\textcolor{MyRed}{S_{i+1}}$ 
& 
$
\begin{array}{lcl}
C[\emptyset] &=& \text{const.} \\
C[\textcolor{MyBlue}{\blacksquare}] 
&=& \overline{\textcolor{MyBlue}{S_i}} \\
C[\textcolor{MyBlue}{\blacksquare}\textcolor{MyRed}{\blacksquare}]
&=&
\overline{\textcolor{MyBlue}{S_i}\textcolor{MyRed}{S_{i+1}}}
\end{array}
$
\\

\hline 

$z_3$ 
& $\text{conv}[{\blacksquare}{\blacksquare}{\blacksquare}]_i =
{W_0}\textcolor{MyBlue}{S_i}
+ {W_1}\textcolor{MyRed}{S_{i+1}}
+ {W_2}\textcolor{MyGreen}{S_{i+2}}$ 
& 
$
\begin{array}{lcl}
C[\emptyset] &=& \text{const.} \\
C[\textcolor{MyBlue}{\blacksquare}] 
&=& \overline{\textcolor{MyBlue}{S_i}} \\
C[\textcolor{MyBlue}{\blacksquare}\textcolor{MyRed}{\blacksquare}]
&=&
\overline{\textcolor{MyBlue}{S_i}\textcolor{MyRed}{S_{i+1}}} \\
C[\textcolor{MyBlue}{\blacksquare}\square\textcolor{MyGreen}{\blacksquare}]
&=&
\overline{\textcolor{MyBlue}{S_i}\textcolor{MyGreen}{S_{i+2}}} \\
C[\textcolor{MyBlue}{\blacksquare}\textcolor{MyRed}{\blacksquare}\textcolor{MyGreen}{\blacksquare}]
&=&
\overline{\textcolor{MyBlue}{S_i}\textcolor{MyRed}{S_{i+1}}\textcolor{MyGreen}{S_{i+2}}}
\end{array}
$
\\

\hline
\end{tabular}
}
\caption{\textbf{Mapping filters to spin correlators.} 
A TetrisCNN branch $\actbottle_k$ computes a linear combination of all correlators $C[P']$ for subpatterns $P'\subseteq P$,
defined from a convolution $\operatorname{conv}[P]$ (which follows from a Boolean Fourier expansion and GAP, see \seclab~\ref{ss:interpretable_Tetris} and \app~\ref{app:boolean_fourier}). 
By additionally promoting sparsity on branch outputs using L1-regularization $\sum_k\lambda_k|\actbottle_k|$, where $\lambda_1<\lambda_2<...<\lambda_K$, TetrisCNN selects \highlight{minimally sized task-relevant correlators}. E.g. if trained with $\actbottle_{1,2,3}$ above and 
$\actbottle_3$ is the only non-vanishing branch output, $
\{\emptyset, 
\textcolor{MyBlue}{\bs},
\textcolor{MyBlue}{\bs}\textcolor{MyRed}{\bs}\}$
are \highlight{task-irrelevant}, since they are included also in 
$\actbottle_2$ which vanished during training {despite weaker penalization}, $\lambda_2<\lambda_3$. Thus one is left with
$\textcolor{MyBlue}{\bs}\square\textcolor{MyGreen}{\bs}$ and/or $ 
\textcolor{MyBlue}{\bs}\textcolor{MyRed}{\bs}\textcolor{MyGreen}{\bs}$.
Note that an unfilled square denotes a masked-out site. For rotation-averaged convolutions $\operatorname{conv}_{\mathrm{rot}}$, the corresponding spin correlators $C_{\mathrm{rot}}$ are likewise rotationally invariant.}
\label{tab:Tetris_correlators}
\end{table} 

\paragraph{Low-dimensional latent space}
A common approach for obtaining interpretable features is to project data onto a \textit{latent space} before making a prediction. By further enforcing it to have specific properties (e.g., a given dimensionality or sparsity), one provides an explicit drive to compress knowledge, and allows one to `disentangle' the neural network's representation \cite{Higgins2017BetaVAE, pmlr-v80-kim18b, Cunningham2024SAE}. 
We employ this strategy in TetrisCNN, although in a non-standard way. That is, we use a composition of mappings:
\begin{equation}
    x \xmapsto{\text { TetrisCNN }} \underbrace{\left\{ \actbottle_k\right\}_{k=1}^K}_{\text {Bottleneck }} \xmapsto{\text { Task NN }} \hat{y},
    \label{eq:tetris_sketch}
\end{equation}
where $\left\{ \actbottle_k\right\}_{k=1}^K$ is the {latent space}, also called {bottleneck}. In typical autoencoder architectures, $x \xmapsto{} z$ is a fully connected neural network. In TetrisCNN, it instead consists of $K$ \textit{parallel} {branches} $x \xmapsto{} \actbottle_k$, with each branch $\actbottle_k=z[P_k]$ defined by \eq~\eqref{eq:branch}. 
This design achieves a second level of interpretability compared to standard autoencoders. Namely, such bottlenecks are only `interpretable' in the sense that they have low dimensionality -- each individual dimension is still a complicated (non-interpretable) function of the data. By contrast, TetrisCNN additionally achieves interpretability of \textit{individual} bottleneck dimensions by virtue of \eq~\eqref{eq:sum_of_correlators}.

\paragraph{Sparsity regularization}
The chosen patterns control the correlations that TetrisCNN can learn. But, as visualized in \tab~\ref{tab:Tetris_correlators}, including branches with larger filters creates redundancy in the correlations picked up by different branches. We break this redundancy by additionally making the bottleneck \textit{sparse} through L1-regularization: 
\begin{equation}\label{eq:L1reg}
\loss_{\mathrm{L1}}=\sum_k \lambda_k |\actbottle_k|,
\end{equation}
where $\lambda_k$ are real-valued hyperparameters. We also call this the \textit{branch penalty}. 
Adding this term to the loss means the network suppresses branches that are not relevant for the task during training. This achieves \textit{automatic feature selection}: branches whose activations remain above the optimization-induced floor (see \app~\ref{app:bottleneck_optimization}) are retained as task-relevant, while unnecessary branches are suppressed.

\paragraph{Penalizing large patterns} The Rashomon effect predicts that multiple models with equally good performance exist for a given dataset \cite{pmlr-v235-rudin24a}, which we also observe in practice (\app~\ref{app:robustness}). 
To break this redundancy, we introduce on top of \eq~\eqref{eq:L1reg} a penalty for `complexity', here defined simply as the cardinality of the pattern $|P|$ (which translates to the \textit{degree} of the corresponding largest correlator, a canonical measure of complexity \cite{schurov2025learning-62d}). We interpolate $\lambda_k$ linearly in log space between $\lambda_{\min}$ and $\lambda_{\max}$ (in the main text equal to $10^{-3}$ and $10^{3}$, respectively) according to $\left|P_k\right|$:
\begin{equation}\label{eq:branch_penalty}
\lambda_k=\lambda_{\min}^{1-\alpha} \lambda_{\max }^\alpha
\end{equation}
where $\alpha \equiv \frac{|P_k|-1}{\max |P_k|-1} \in[0,1]$.
This means that branches with larger patterns incur a higher cost and are thus {suppressed} unless they provide additional predictive power compared to branches with smaller patterns. In this way, since \eq~\eqref{eq:L1reg} makes the bottleneck sparse, \textit{TetrisCNN is biased toward the smallest pattern that provides predictive information sufficient to solve a task.} We will refer to suppressed branches as `inactive' or `deactivated', and the remaining branches as `active'. In practice, we find that learning rate scheduling (\app~\ref{app:lr_scheduling}) helps accentuate the difference between active and inactive branches. 

\paragraph{Task network}
The innovative part of TetrisCNN relies on the interpretability of its bottleneck. These bottleneck activations are in the end input to a standard black-box fully connected network ($\{\actbottle_k\} \xmapsto{} \hat y$), which we refer to as a task network (grey layer in \fig~\ref{fig:intro}). The degree to which the task network can be interpreted depends on the task. In the classification task, we treat the task network as a classifier operating in the bottleneck space, thereby yielding interpretable decision boundaries, as presented in the text below. In more complex settings, one can resort to symbolic regression (see \app~\ref{app:SR}, also for criticisms of the method), which becomes cheap due to the bottleneck's interpretability and sparsity.

\begin{figure*}[t]
    \centering
    \includegraphics[width=0.98\textwidth]{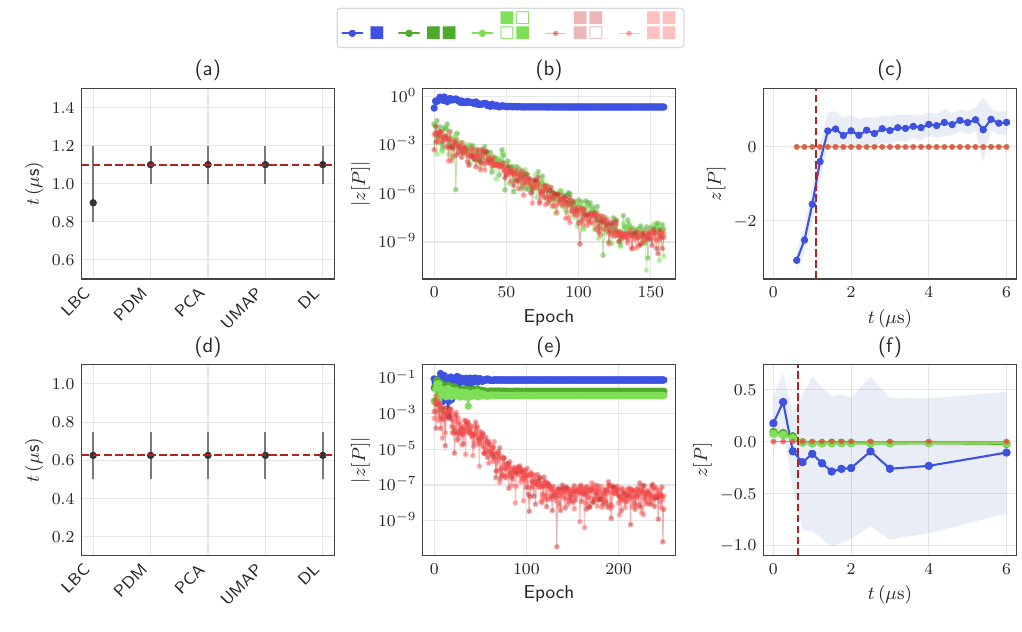}
    \caption{\figcap{Data-driven transition detection and interpretable
    classification from experimental snapshots} for (a)--(c) Ising
    and (d)--(f) XY. 
    (a,d) Transition locations estimated with established unsupervised methods (\app~\ref{app:unsupervised_comparison}). Red dashed lines mark the location used to define supervised labels. For Ising, the detected change is the polarization crossover rather than the later antiferromagnetic phase transition. (b,e) Evolution of rotationally invariant TetrisCNN branch activations during training, showing sparsity-induced suppression of task-irrelevant branches. (c,f) Activations of the surviving branches across the experimental sweeps. Unfilled squares denote masked-out sites.
}
\label{fig:interpretable_detection}
\end{figure*}

\paragraph{TetrisCNN hyperparameters}
TetrisCNN has two main hyperparameters. One is the maximum branch penalty $\lambda_{\max}$. Increasing $\lambda_{\max}$ penalizes larger-filter branches more strongly and therefore favors solutions based on fewer and lower-order correlators. In \app~\ref{app:robustness}, we show that our results are robust to the specific choice of $\lambda_{\max}$, once a minimum threshold is reached. The second architectural hyperparameter is the choice of the set of patterns $\{P_k\}$. 
Here we take all subpatterns within a given $2 \times 2$ receptive field, $P_k\subseteq \sus{\bs\bs\\\bs\bs}$. We justify this in \app~\ref{app:search_for_scale}, where another TetrisCNN run with larger filter sizes ($1\times1$, $2\times2$, $4\times4$, $8\times8$) finds that 
scales beyond $2 \times 2$ are irrelevant for the tasks at hand. This procedure can also be viewed as a data-driven search for the spatial scale of the correlators relevant to the task. 
Moreover, we impose the $C_4$ rotational symmetry of the square lattice, so each filter is applied in all distinct $90^\circ$ rotations of its pattern, and the resulting feature maps are averaged. Consequently, each branch is invariant under rotations of its input pattern. In \app~\ref{app:rotational_invariance} we show that rotationally invariant and unconstrained variants of TetrisCNN give similar results.
For further architectural details of TetrisCNN and information on the hyperparameters used, we refer the reader to \app~\ref{app:archi_hyperparams}. The optimization details are in \app~\ref{app:bottleneck_optimization}.

\section{Results}\label{sec:results}

\subsection{Sparse description of the identified transitions}

\paragraph{Unsupervised detection of transitions across the experimental sweeps}
We first identify the locations of transitions in the two experimental datasets using several established unsupervised learning approaches, namely learning by confusion (LBC), the prediction-divergence method (PDM), principal component analysis (PCA), Uniform Manifold Approximation and Projection (UMAP), and distance learning (DL). Here and throughout, we use \emph{transition} broadly to encompass both phase transitions and crossovers. The resulting location estimates are shown in \fig~\ref{fig:interpretable_detection}(a,d) and discussed in detail in \app~\ref{app:unsupervised_comparison}. The transition location with the best agreement among the methods is $t=1.1\,\mu\mathrm{s}$ for the Ising data, while the estimated transition location for the XY data is $t=0.625\,\mu\mathrm{s}$. These locations are marked by red dashed lines in panels (a),(c)--(d), and (f) of \fig~\ref{fig:interpretable_detection}. For all methods, the error bars span the acquisition times flanking the inferred transition. For LBC, they also reflect variation in the inferred transition location across runs. As discussed in \seclab~\ref{sec:discussion}, the change detected in the Ising data coincides with the loss of the initial $Z$-polarization rather than the later emergence of antiferromagnetic order.

\paragraph{Supervised training of TetrisCNN}
Having identified the relevant transition in each dataset, we construct a supervised classification task by assigning different labels to snapshots from opposite sides of the identified location and train TetrisCNN on
this task. The purpose of this step is not to improve the location estimate, but to reveal which local  observables are most relevant for distinguishing the two classes.
TetrisCNN reaches validation accuracies of $99.14\%$ and $93.54\%$ for the Ising and XY datasets, respectively. In both cases, most misclassifications occur near the identified transition (see \fig~\ref{fig:X_Z_XZ_errors}(a),(b) in \app~\ref{app:separate_X_Z_training}), partly due to overlapping spin configurations on opposite sides of the transition.

\paragraph{Evolution of TetrisCNN branch activations} \Figs~\ref{fig:interpretable_detection}(b,e) show the evolution of branch activations during training. Due to the sparsity regularization, branches that are not required for the classification task gradually `deactivate'. For the Ising dataset, a single branch $z[\bs]$ remains active after training, while all other branches are suppressed by several orders of magnitude. In contrast, multiple branches remain active for the XY dataset, indicating that a richer set of descriptors is required to distinguish the corresponding classes.

The activations of the branches across the experimental sweep are shown in \fig~\ref{fig:interpretable_detection}(c,f). In both datasets, the selected descriptors change sharply around the identified transition.
In \app~\ref{app:robustness}, we show that the identified branches are robust to the choice of sparsity strength.
Having established that only a small subset of branches is required for classification, we next identify the physical observables encoded by the active branches.

\subsection{Mapping branch activations to spin correlators} 

\paragraph{Boolean Fourier expansion} A key advantage of TetrisCNN is that its interpretable architecture permits a direct interpretation of its latent representation in terms of physically meaningful spin correlators. To identify the observables encoded by the active branches, we express each branch activation as a multilinear expansion in the correlators associated with the corresponding filter pattern using the Boolean Fourier expansion, as described in \seclab~\ref{ss:interpretable_Tetris}; see \eq~\eqref{eq:sum_of_correlators}.

\begin{figure}[t]
    \centering
    \includegraphics[width=0.98\columnwidth]{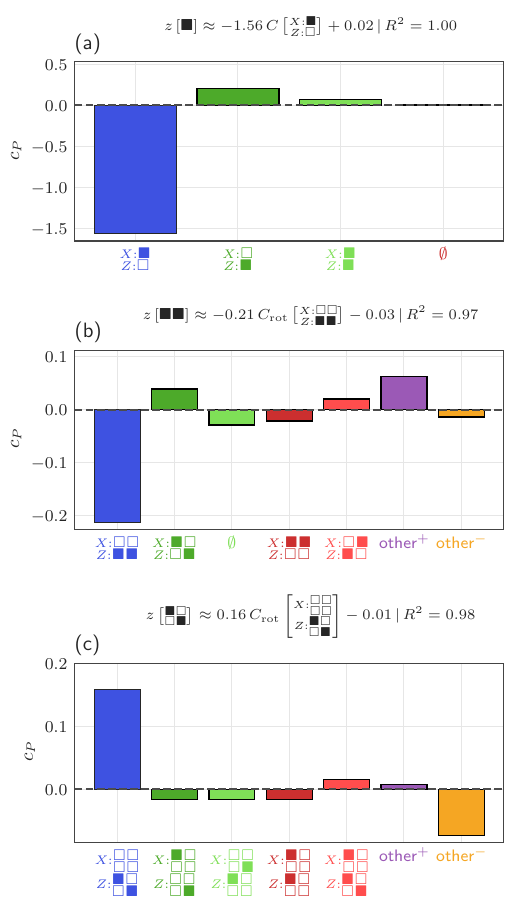}
    \caption{\figcap{Mapping from branch activations to correlators} via multilinear regression for three active branches in TetrisCNN trained on the XY dataset as in \fig~\ref{fig:interpretable_detection}(e)-(f), enabled by the Boolean Fourier expansion, \eq~\eqref{eq:sum_of_correlators}. We need only to consider correlators $C[P'], P'\subseteq P$ defined by patterns $P'$ that are contained in the filter pattern $P$ (see \tab~\ref{tab:Tetris_correlators}). Note that with all $C[P'], P'\subseteq P$, the fit is perfect ($R^2=1$). Small positive (negative) contributions are summed into ``other+(--)''. The complete decomposition is shown in \fig~\ref{fig:XY_multilinear_regression_full}.}\label{fig:XY_multilinear_regression}
\end{figure}

\paragraph{Magnetization in the Ising dataset} 
For the Ising dataset, as shown in \fig~\ref{fig:interpretable_detection}(b), only the $z[\bs]$ branch remains active after training. From \tab~\ref{tab:Tetris_correlators}, we see that the only correlators that a branch with a filter $\bs$ can compute are 
$C[\bs]$, so in this case $Z$-basis magnetization, $C^Z[\bs] = \overline{S_i^Z}$. A linear regression fit gives 
\begin{equation}\label{eq:Ising_activation}
    z[\bs] = -4.626 \, C^Z[\bs] + 0.954\,. 
\end{equation}
Thus, despite having access to a large collection of candidate correlators, TetrisCNN automatically identifies a 1D description of the phase classification problem.

\paragraph{Three correlators are needed for XY data}
In the XY case, the network uses a richer set of correlators. As shown in \fig~\ref{fig:interpretable_detection}(e), not only the $z[\bs]$ branch but also the $z[\bs \bs]$ and $z[\sus{\bs \square \\ \square \bs}]$ branches remain active, indicating that the task cannot be solved with high accuracy with a single local observable. 
In \fig~\ref{fig:XY_multilinear_regression}, the multilinear regression reveals that the $z[\bs]$ branch is dominated by the $X$-basis magnetization $C^X[\bs]$, while the $z[\bs \bs]$ and $z[\sus{\bs \square \\ \square \bs}]$ branches compute rotationally invariant $Z$-spin correlators, \textit{i.e.}, $C_{\mathrm{rot}}^Z[\bs \bs]$ and $C_{\mathrm{rot}}^Z[\sus{\bs \square \\ \square \bs}]$, respectively. The linear fits of each activation to its dominant spin correlator, given in full in \fig~\ref{fig:XY_multilinear_regression}(a)-(c), provide an almost exact description, with $R^2=1.00$, $0.97$, and $0.98$, respectively.

\paragraph{Variance of the $z[\bs]$ branch}
Interestingly, the activation of the $z[\bs]$ branch in \fig~\ref{fig:interpretable_detection}(f) has much larger variance than other branches, particularly after the phase transition. We will resolve this riddle in the next section once we identify the function of the spin correlator used by the classifier through $z[\bs]$.

Taken together, the Ising and XY results demonstrate that TetrisCNN not only identifies a sparse set of descriptors, but also reveals their physical interpretation in terms of experimentally measurable spin correlators.

\subsection{Decision boundaries in interpretable latent space}

\paragraph{Bottleneck space is interpretable}
So far, we have identified which spin correlators the network computes with its active branches to make the predictions. Now we can take a step further and understand how the network will classify a new snapshot by identifying the symbolic formula for its decision boundary.
We achieve this by analyzing the network predictions directly in the latent space spanned by the branch activations. Crucially, this low-dimensional latent space is no longer an abstract embedding, as in, for example, autoencoders \cite{wetzel_unsupervised_2017}: each axis has a known physical meaning in terms of spin correlators.

\begin{figure}[t]
    \centering
     \includegraphics[width=0.98\columnwidth]{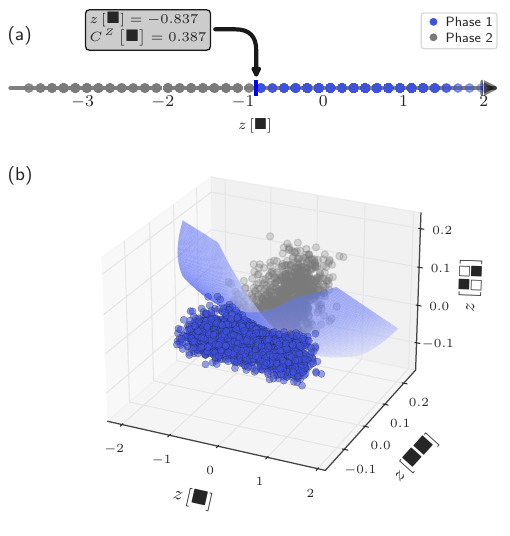}
    \caption{\figcap{Decision boundaries in the interpretable latent space of TetrisCNN} trained on the Ising (a) and XY (b) datasets. Each point represents one experimental snapshot, positioned according to its latent activations and colored by the phase label predicted by the classifier. (a) In the Ising case, only a single branch remains active, yielding a 1D latent space in which the classifier acts by thresholding the $Z$-basis magnetization. (b) In the XY case, the latent space is spanned by the three active branches identified in \fig~\ref{fig:interpretable_detection}(e)-(f). The displayed surface is a quadratic approximation to the learned decision boundary, fitted on the training data, and reproduces the network's predicted labels on the validation data with $99.84$\% accuracy. Because each latent coordinate corresponds to a known spin correlator (\fig~\ref{fig:XY_multilinear_regression}), both classification rules can be expressed directly in terms of physically meaningful observables, cf. \eq~\eqref{eq:decision_boundary_in_activations} $\rightarrow$ \eq~\eqref{eq:decision_boundary_in_correlators}.
    }
\label{fig:decision_boundaries}
\end{figure}

\paragraph{The Ising bottleneck is 1D}
Disregarding the deactivated branches, we plot the activations of the only remaining active branch $z[\bs]$ for all snapshots from the validation set in \fig~\ref{fig:decision_boundaries}(a) and color code them by their predicted label. This reveals the full decision rule: the decision boundary is a single point at 
$z[\bs]=-0.837$,
which, as we know from \eq~\ref{eq:Ising_activation}, corresponds to a $Z$-magnetization threshold of 
$C^Z[\bs] = \overline{S_i^Z} = 0.387$.

\paragraph{The XY bottleneck is higher-dimensional}  
Again, we plot all snapshots in the space of active branches, which for XY is three-dimensional (3D) and spanned by
$z[\bs]$, $z[\bs \bs]$, and $z[\sus{\bs \square \\ \square \bs}].$ 
Here, the decision boundary is a surface. We approximate it by fitting a polynomial surrogate to the TetrisCNN predictions, as detailed in \app~\ref{app:dec_bound_log_reg_fit}. We find that an almost perfect ($99.84\%$) approximation is quadratic:
\begin{equation}\label{eq:decision_boundary_in_activations}
    0.05 z[\bs]^{2} - 0.03 z[\bs] - 0.9 z[\bs \bs] - z[\sus{\bs \square \\ \square \bs}] + 0.03 = 0\,.
\end{equation}
Excitingly, we can use the mapping from activations to spin correlators in \fig~\ref{fig:XY_multilinear_regression}, and express this decision boundary in terms of spin correlators:
\begin{equation}\label{eq:decision_boundary_in_correlators}
\begin{aligned}
    0.68 C^{X}[\bs]^2 + 0.21 C^{X}[\bs] \\
    {}+ C^{Z}_{\mathrm{rot}}[\bs\bs] - 0.81 C^{Z}_{\mathrm{rot}}
    [\sus{\bs\square \\ \square\bs}]
    + 0.35 &= 0\,.
\end{aligned}
\end{equation}
which yields the final result: an explanation of the network's classification rule in terms of physically relevant observables. The identified formula is robust across random initializations and random re-pairings of the $X$- and $Z$-basis snapshots, and we provide robustness tests and fitting details in \app~\ref{app:xz_pairing_ablation} and \app~\ref{app:dec_bound_log_reg_fit}, respectively. The identified quadratic contribution
shows that identifying the relevant observables alone is not sufficient to explain the classifier. The task network learns a nontrivial function of these observables (such as $z[\bs]^2$), which leads beyond a linear classification in the latent space. In this case, 
the quadratic contribution from $C^X[\bs]$ provides an improvement, especially at the phase transition where it reduces the disagreement with the TetrisCNN predictions from $3\%$ to zero, as shown in \app~\ref{app:plane_vs_surface}. 

Supporting this interpretation, TetrisCNNs trained separately on the two measurement bases (\app~\ref{app:separate_X_Z_training}) recover a quadratic function of the $X$-magnetization from $X$-basis snapshots, while the $Z$-basis network recovers the same linear function of the two $Z$-spin correlators identified above. The latter provides substantially stronger single-snapshot discrimination, reaching $91.19\%$ validation accuracy compared with $70.28\%$ for the $X$-only network. Thus, TetrisCNN combines the stronger overall single-snapshot classification signal from the $Z$-basis correlators with the physically expected squared $X$ magnetization, whose contribution is particularly important near the identified phase transition.

\paragraph{Order parameter and the branch variance} 
Remarkably, this quadratic term $C^{X}[\bs]^2$ closely relates to the order parameter of the phase transition in the XY dataset identified by \reflab~\cite{Chen2023qsimXY}; the in-plane (ferromagnetic) magnetization squared:
\begin{equation}
    (m^X)_c^2=\frac{1}{N^2} \sum_{i, j} \widetilde C_{i, j}^X
\label{eq:mX2}
\end{equation}
where
$\widetilde C_{i, j}^X=\left\langle\sigma_i^X \sigma_j^X\right\rangle-\left\langle\sigma_i^X\right\rangle\left\langle\sigma_j^X\right\rangle$ (not to be confused with $C[P]$ from \eq~\eqref{eq:C(P)}). Rewriting, we obtain:
\begin{equation}
(m^X)^2_{\mathrm{c}} =  \left\langle C^X[\bs]^2\right\rangle-\left\langle C^X[\bs]\right\rangle^2\,,
\end{equation}
where TetrisCNN bases its decision boundary on the single-snapshot quantity $C^X[\bs]^2$, whose average over snapshots gives the first term of the connected order parameter, $\left\langle C^X[\bs]^2\right\rangle$.
Finally, we can also explain the variance of the $z[\bs]$ branch in the XY dataset. The shaded region in \fig~\ref{fig:interpretable_detection}(f) is:
\[
\sqrt{\left\langle
\left(z-\langle z\rangle\right)^2
\right\rangle}
=
\sqrt{\langle z^2\rangle-\langle z\rangle^2}\,.
\]
Here, $z=z[\bs]$. Using the fit in \fig~\ref{fig:XY_multilinear_regression}(a), $z[\bs]\approx C^X[\bs]$. Thus, the squared width of the $z[\bs]$ activation distribution is approximately the variance of $C^X[\bs]$,
which is directly related to the in-plane ferromagnetic order parameter $(m^X)^2$ defined in \eq~\eqref{eq:mX2}.

\section{Discussion}\label{sec:discussion}

\paragraph{Polarization crossover in the Ising data}
All machine-learning approaches considered in this work, including learning by confusion, prediction divergence, principal component analysis, UMAP, and distance learning, consistently identify a transition around $t\approx1.1\,\mu\mathrm{s}$. The interpretability of TetrisCNN reveals that the corresponding classification signal is carried by the uniform $Z$-magnetization. This signal does not correspond to the later onset of antiferromagnetic order. Instead, it reflects the smooth crossover that occurs as the increasing detuning $\longitudinal$ drives the system out of the nearly fully $Z$-polarized state and into a more strongly mixed paramagnetic regime while the transverse field $\transverse$ remains finite.

\paragraph{Antiferromagnetic transition in the Ising data}
The later paramagnetic-to-antiferromagnetic quantum phase transition is signaled by the growth of the staggered magnetization in
\fig~\ref{fig:quantum_data}(b). On the finite experimental arrays, the transition is rounded, and the staggered magnetization increases continuously rather than displaying the sharp nonanalytic behavior expected in the thermodynamic limit. The system-size comparison reported in \reflab~\cite{Scholl2021qsimIsing} shows that this increase becomes sharper for larger arrays, consistent with finite-size rounding, while the finite ramp duration and experimental imperfections introduce additional broadening.

For reference, ground-state density-matrix-renormalization-group calculations in  \reflab~\cite{Scholl2021qsimIsing} locate the inflection point of the staggered magnetization near $t\approx3.2\,\mu\mathrm{s}$ for a $6\times6$ system and $t\approx2.6\,\mu\mathrm{s}$ for a $10\times10$ system. Interestingly, none of the automated methods considered here detects this later transition, even though its signatures are visible in several correlators, including $C_{\mathrm{rot}}^Z[\bs\bs]$. In contrast, a recent graph-theoretic data-driven approach reports a transition location consistent with the
emergence of antiferromagnetic order \cite{mendes-santos2024wave-function}. Understanding why this broad
class of methods preferentially detects the earlier polarization crossover and whether this preference is related to the different sharpness of the two changes remains an open question.

\paragraph{The outcome of interpretation - or what we can learn}
More generally, the observables identified by TetrisCNN should be interpreted as those the network uses to solve a given learning task, rather than as straightforward order parameters or observables most fundamental to the physics of interest. The XY dataset provides an instructive example. Here, TetrisCNN identifies the quadratic contribution of the $X$-basis magnetization to the decision boundary, in close agreement with the established order parameter of \reflab~\cite{Chen2023qsimXY}. However, this quantity alone does not provide sufficient information to reliably classify individual snapshots: a TetrisCNN trained only on $X$-basis snapshots reaches substantially lower accuracy than one trained on $Z$-basis snapshots (\app~\ref{app:separate_X_Z_training}). The network therefore supplements the physically expected squared $X$ magnetization with two less conventional $Z$-basis correlation functions, $C^Z[\bs \bs]$ and $C^Z[\sus{\bs \square \\ \square \bs}]$, which provide a much stronger single-snapshot classification signal. A linear classifier built solely from these two correlators reproduces approximately 96\% of the network predictions (see \app~\ref{app:fitting_decision_boundary}), even though neither quantity is conventionally regarded as an order parameter of the transition. 
This illustrates an important distinction between identifying observables conventionally associated with the phase transition and identifying those most useful for a particular learning task. The latter need not coincide with the order parameter or with observables directly tied to the underlying symmetry breaking. For example, \reflab~\cite{Chen2023qsimXY} notes that systematic measurement errors lead to a small but nonzero value of $\langle m^X \rangle$ in the studied dataset. If such a feature were sufficiently predictive of the target labels, TetrisCNN could exploit it. A standard black-box neural network could exploit the same artifact while leaving its role hidden, so high classification accuracy alone could be mistaken for evidence that the network had learned the underlying physics. By contrast, TetrisCNN exposes the observables used by the classifier, allowing them to be scrutinized against theoretical expectations and known experimental imperfections. Physical insight therefore remains necessary to determine whether the identified observables reflect the conventional order parameter or symmetry-breaking physics, other interesting physical structure in the data, or experimental artifacts.

\paragraph{Task dependence of the learned correlators} Moreover, the correlators selected by TetrisCNN depend on the learning objective. In this work, we focused on supervised phase classification, which favors observables that change rapidly across the transition and therefore provide maximal discriminative power between phases. As a consequence, the identified descriptors often resemble conventional phase indicators. However, when the learning objective changes, the network may prefer a different set of correlators better suited to the task at hand. 
For example, when TetrisCNN is used within prediction-divergence method to regress the experimental tuning parameters from snapshots, it selects correlators informative of continuous variation of the target along the experimental sweep (see \app~\ref{app:task_dependence}). 

\paragraph{Rashomon effect} Another interesting observation is that the sparsity regularization of TetrisCNN acts primarily as a model-selection mechanism rather than as a performance-enhancing regularizer. As shown in \app~\ref{app:robustness}, many distinct internal representations achieve nearly identical classification accuracies. Increasing the sparsity regularization progressively removes more complex branches while leaving the predictive performance essentially unchanged, selecting a simple representative from a large family of functionally equivalent solutions. The phenomenon of multiple equally good predictive models existing for the same dataset is known as the Rashomon effect \cite{pmlr-v235-rudin24a} (named after Akira Kurosawa's movie \textit {Rashomon}, in which several characters give mutually incompatible accounts of the same event, each plausible from their perspective). It is especially relevant to experimental datasets, as the number of such equally good predictive models tends to increase with data noise. Fortunately for interpretability enthusiasts such as ourselves, the large number of such models also correlates well with the existence of simple yet accurate models \cite{semenova2022existence-a79, semenova2023a}.
From a physics perspective, it suggests that phase transitions can often be detected in experimental data through various observables. Indeed, several correlators in \fig~\ref{fig:quantum_data}(b),(d) can be used to distinguish the phases equally well, explaining why different branch combinations can yield nearly identical transition estimates. TetrisCNN, therefore, does not identify a unique set of descriptors relevant to phase transition detection, but rather a particularly simple and interpretable one.

\section{Related work}\label{sec:related_work}

Automatic detection of phase transitions and the corresponding order parameters has been a widely recognized goal of machine learning applied to (quantum) physics since 2017~\cite{wetzel2017orderparams}. Rather than replacing existing methods, we view TetrisCNN as complementing the growing toolbox of interpretable approaches.

One of the most successful solutions has been to pair neural networks with a renormalization group procedure based on mutual information \cite{Koch-Janusz2018, gkmen2021symmetries-a1f, gkmen2021statistical-c8c, gordon2021relevance-893}, which thrives in the regime of large classical systems and can even be used to study inhomogeneous and biological systems \cite{gkmen2024compression-f99}.
Tensorial-kernel support vector machines \cite{Greitemann19probing, Liu19interpretable, Greitemann19identification, Liu2021, rao2021inferring-b02, sadoune2023unsupinterpet, sadoune2025human-machine-a0b, sadoune2026learning-c10}, thanks to their impressive interpretability, have led to one of the few scientific discoveries with machine learning in physics \cite{Liu2021} but at the expense of substantial computational cost and the need to handcraft candidate correlators. Correlator CNN and its extensions \cite{Miles2021CCNN, schlomer2023fluctuationinterpret, gong2024c3nn-6ea, suresh2025interpretable-add} learn physically meaningful filters and can even tackle long-range order \cite{schlomer2023fluctuationinterpret} but require visual inspection to interpret them.
Probabilistic variational autoencoders can successfully recover phase structure and candidate phenomena from unlabeled quantum data, including noisy experimental snapshots and nonequilibrium protocols, by learning their compact representation, but interpreting this representation and extracting explicit descriptors still require post-hoc analysis via symbolic regression or substantial physical guidance \cite{schoulepnikoff2025interpretable-3a6, moller2026learning-648, schoulepnikoff2026discovering-e9b}.
Last but not least, one can also resort to exhaustive searches over candidate correlators \cite{verdel24datadriven}.

TetrisCNN combines several of these desirable properties and offers new ones. It evaluates many candidate correlators simultaneously, automatically selects a sparse subset via regularization, and expresses the resulting classifier directly as symbolic formulas that describe decision boundaries acting on physically meaningful observables. In contrast to post-hoc symbolic-regression approaches \cite{schoulepnikoff2026discovering-e9b}, whose recovered expressions depend on the user-defined search space and can extrapolate poorly when fitted to the full network output (see \app~\ref{app:SR}), the physical meaning of the TetrisCNN latent coordinates follows directly from the architecture and Boolean-Fourier expansion. The remaining low-dimensional decision rule can then be approximated with a controlled polynomial surrogate. At the same time, the current implementation loses its full interpretability when detecting long-range correlations, allowing only the smallest task-relevant scale to be determined.

\section{Conclusion and outlook}\label{sec:conclusion}

Physical models are often sparse descriptions of complex phenomena \cite{roberts2021why-960}. In the context of phases of matter, order parameters provide a canonical example of this principle: they provide sparse representations of many-body data sufficient to characterize a phase transition. In this work, we translated this principle into an interpretable convolutional neural-network architecture called TetrisCNN, which uses sparsity as an inductive bias to identify a small set of spin correlators sufficient for solving a given learning task.
TetrisCNN combines parallel convolutional branches with filters of different sizes and shapes and promotes sparse bottleneck representations through L1-regularization, which favors branches with smaller filters over larger ones. Thanks to the exact mapping between the bottleneck activations and spin correlators, both the latent representation and the learned decision boundaries can be expressed as symbolic formulas in terms of physically meaningful observables. Applied to experimental Rydberg snapshot data for the Ising and XY models, this approach reveals which spin correlators are used to make the prediction and how they are combined to distinguish phases of matter.

In the systems studied here, TetrisCNN uncovered crossover and phase indicators that agree with existing theoretical intuition. In the transverse-field Ising dataset, the network bases its prediction on the $Z$-magnetization, revealing that it detects the crossover out of the trivial polarized state rather than the later emergence of antiferromagnetic order. In the XY dataset, the network identifies a sparse combination of nearest-neighbor and diagonal two-site $Z$ correlators, together with a quadratic contribution from the $X$-magnetization, closely related to the established in-plane ferromagnetic order parameter. More generally, the ability to express the network predictions in terms of physical observables enables them to be scrutinized, validated, or challenged by physicists.

We view TetrisCNN as a step toward machine-assisted discovery. As quantum simulators continue to explore increasingly complex quantum systems, interpretable networks can quickly provide sparse representations of data and help in understanding their new behaviors. To make this happen with TetrisCNN, the next step is to move beyond local order parameters by enriching the branch dictionary to capture nonlocal, string, or other more exotic correlators. Another challenge is to move beyond regular lattices, where geometry itself may generate unfamiliar phases of matter and where standard order-parameter intuition is less developed. More broadly, the TetrisCNN architecture is independent of the learning objective. In this work, we focused on supervised phase classification, but the same interpretable architecture can be combined with a wide range of machine-learning paradigms, including distance-learning objectives, prediction-divergence methods, neural quantum states \cite{Valenti2021}, and generative models \cite{Casert21PRR, fitzek2025rydberggpt-c7c, moller2026learning-648}. This flexibility enables a broad class of machine-learning approaches to become interpretable, extending their role from predictive tools to sources of physical insight into complex many-body systems. Ultimately, we envision machine learning evolving from a tool that makes predictions into one that explains them in the language of physics, enabling an iterative dialogue between experiment, theory, and data-driven models.

\section*{Data and code availability}
The data that support the findings of this article and the code used in this study are publicly available at \reflab~\cite{our_github_2026}.

\begin{acknowledgments}
We thank Gorka Muñoz-Gil for useful discussions.
BvZ and AD acknowledge support from the
Dutch National Growth Fund (NGF), as part of the Quantum Delta NL programme and the Top Talent award.
The Flatiron Institute is a division of the Simons Foundation. This research was supported in part by grant no. NSF PHY-2309135 to the Kavli Institute for Theoretical Physics (KITP). This work was performed using the ALICE compute resources provided by Leiden University. 
\end{acknowledgments}

\bibliographystyle{apsrev4-2.bst}
\bibliography{bibliography}

\vspace{2em}

\begingroup
\setlength{\parindent}{0pt}
\setlength{\parskip}{0pt}
\noindent\textbf{Contents of the appendices}\par
\begin{footnotesize}
\vspace{4pt}

\hyperref[app:data_preparation]{\textbf{\ref{app:data_preparation}.} Data preparation}\\[1pt]
\hspace*{1.1em}\hyperref[app:snapshot_imbalance]{\ref{app:snapshot_imbalance}. Snapshot imbalance and data splitting}\\[1pt]
\hspace*{1.1em}\hyperref[app:combining_bases]{\ref{app:combining_bases}. Combining two measurement bases}\\[5pt]

\hyperref[app:correlators]{\textbf{\ref{app:correlators}.} Correlators}\\[1pt]
\hspace*{1.1em}\hyperref[app:boolean_fourier]{\ref{app:boolean_fourier}. Boolean function theory}\\[1pt]
\hspace*{1.1em}\hyperref[app:counting_correlators]{\ref{app:counting_correlators}. Counting patterns}\\[5pt]

\hyperref[app:archi_hyperparams]{\textbf{\ref{app:archi_hyperparams}.} Details of TetrisCNN architecture and hyperparameters}\\[5pt]

\hyperref[app:bottleneck_optimization]{\textbf{\ref{app:bottleneck_optimization}.} Separation of active and deactivated branches by learning-rate annealing}\\[1pt]
\hspace*{1.1em}\hyperref[app:noise_floor]{\ref{app:noise_floor}. Origin of the activation floor}\\[1pt]
\hspace*{1.1em}\hyperref[app:survival_threshold]{\ref{app:survival_threshold}. Active versus deactivated branches}\\[1pt]
\hspace*{1.1em}\hyperref[app:lr_scheduling]{\ref{app:lr_scheduling}. Sharpening sparsity with learning-rate annealing}\\[5pt]

\hyperref[app:unsupervised_comparison]{\textbf{\ref{app:unsupervised_comparison}.} Unsupervised detection of transitions in the experimental sweeps: method comparison}\\[1pt]
\hspace*{1.1em}\hyperref[app:lbc]{\ref{app:lbc}. Learning by confusion}\\[1pt]
\hspace*{1.1em}\hyperref[app:pdm]{\ref{app:pdm}. Prediction-divergence method}\\[1pt]
\hspace*{1.1em}\hyperref[app:dimred]{\ref{app:dimred}. Dimensionality reduction and clustering}\\[1pt]
\hspace*{1.1em}\hyperref[app:dl]{\ref{app:dl}. Distance learning}\\[1pt]
\hspace*{1.1em}\hyperref[app:comparison]{\ref{app:comparison}. Comparison of transition estimates}\\[5pt]

\hyperref[app:supporting_material]{\textbf{\ref{app:supporting_material}.} Supporting material}\\[1pt]
\hspace*{1.1em}\hyperref[app:robustness]{\ref{app:robustness}. Robustness of the learned descriptors to regularization}\\[1pt]
\hspace*{1.1em}\hyperref[app:search_for_scale]{\ref{app:search_for_scale}. Searching for the relevant scale}\\[1pt]
\hspace*{1.1em}\hyperref[app:rotational_invariance]{\ref{app:rotational_invariance}. Rotationally invariant and unconstrained TetrisCNN}\\[1pt]
\hspace*{1.1em}\hyperref[app:xz_pairing_ablation]{\ref{app:xz_pairing_ablation}. Robustness to random $X$/$Z$ pairings}\\[1pt]
\hspace*{1.1em}\hyperref[app:separate_X_Z_training]{\ref{app:separate_X_Z_training}. Contribution of the $X$ and $Z$ bases to classification}\\[1pt]
\hspace*{1.1em}\hyperref[app:task_dependence]{\ref{app:task_dependence}. Task dependence of the learned descriptors}\\[1pt]
\hspace*{1.1em}\hyperref[app:full_map_to_corrs]{\ref{app:full_map_to_corrs}. Complete mapping from filters to correlators}\\[5pt]

\hyperref[app:multivariate_R2]{\textbf{\ref{app:multivariate_R2}.} $R$-squared for multivariate and aggregate output}\\[5pt]

\hyperref[app:fitting_decision_boundary]{\textbf{\ref{app:fitting_decision_boundary}.} Fitting the decision boundary to the classifier predictions in the bottleneck space}\\[1pt]
\hspace*{1.1em}\hyperref[app:dec_bound_log_reg_fit]{\ref{app:dec_bound_log_reg_fit}. Fitting the decision boundary}\\[1pt]
\hspace*{1.1em}\hyperref[app:plane_vs_surface]{\ref{app:plane_vs_surface}. Decision boundary: plane vs.\ quadratic surface}\\[5pt]

\hyperref[app:SR]{\textbf{\ref{app:SR}.} Limitations of symbolic regression}

\end{footnotesize}
\endgroup

\newpage

\appendix

\newpage
\section{Data preparation}\label{app:data_preparation}
\subsection{Snapshot imbalance across the sweep and data splitting}\label{app:snapshot_imbalance}

\begin{figure}[b]
    \centering
    \includegraphics[width=0.98\columnwidth]{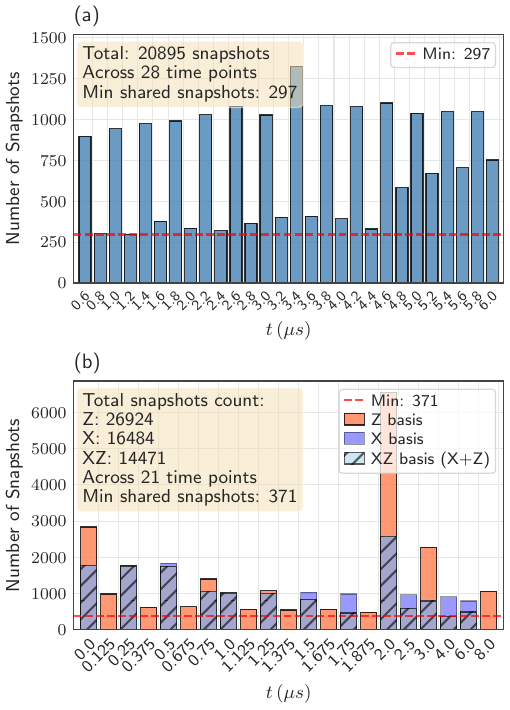}
    \caption{\figcap{Snapshot counts vary by up to an order of magnitude across the sweep.} Distribution of the acquired snapshots over the parameter sweep steps for (a) the $8\times8$ transverse-field Ising model and (b) the $6\times7$ XY model. Red dashed lines indicate the minimum snapshot count ($297$ for the Ising dataset and $371$ for the XY dataset).}
    \label{fig:number_of_snapshots}
\end{figure}

A characteristic challenge of working with raw projective measurements from quantum simulators, as opposed to numerically simulated ground states, is the highly non-uniform distribution of snapshots acquired across the experimental sweep. As shown in \fig~\ref{fig:number_of_snapshots}, the number of snapshots varies by up to an order of magnitude along the experimental sweep. Throughout this work, we therefore use a stratified $70/30$ train-validation split performed independently at each sampling point, ensuring that all regions of the sweep are represented in both sets.

We verified that our results are insensitive to this choice by comparing the stratified split with three alternatives: a global random split, a stratified split with the number of snapshots per sampling point capped at 500, and training with an inverse-frequency-weighted loss. Across these choices, we observe only modest differences in validation accuracy, within $2$ percentage points. We therefore retain the stratified split as the simplest approach that preserves all available experimental snapshots without modifying their contribution to the loss.

\begin{figure}[t]
    \centering
    \includegraphics[width=\linewidth]{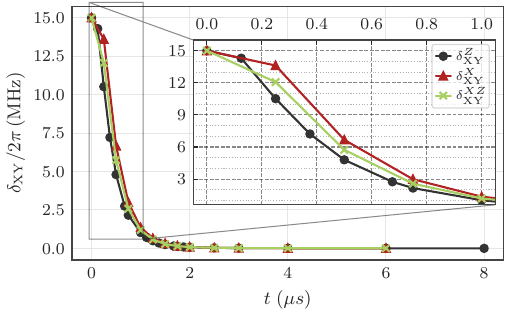}
    \caption{\figcap{The two measurement bases are acquired in separate sweeps whose control fields do not coincide.} Realized staggered detuning $\deltaxy$ across the experimental sweeps for the independent $X$-basis and $Z$-basis measurements of the XY system. The inset highlights the discrepancy during the initial rapid ramp-down. The effective tuning parameter $\deltaxy^{XZ}$ (green curve) is the arithmetic mean of the two physical sweeps, used when combining snapshots from both measurement bases.}
    \label{fig:app:XY_delta_avg}
\end{figure}

\subsection{Combining two measurement bases}\label{app:combining_bases}

Projective measurements are destructive, so the $X$- and $Z$-basis snapshots of the XY dataset originate from separate experimental realizations performed under nominally identical conditions. Combining them therefore requires addressing two differences between the datasets: small discrepancies in the realized control parameters and the absence of a ground-truth pairing between individual snapshots.

\paragraph{Differing Hamiltonian parameters.}
As shown in \fig~\ref{fig:app:XY_delta_avg}, the staggered detuning $\delta_{\mathrm{XY}}$ of the XY experiment is similar between the two measurement sweeps except during the initial rapid ramp-down, where the discrepancy reaches approximately $\sim 3$~MHz. For the combined $XZ$ dataset, we therefore define the effective tuning parameter as the arithmetic mean, $\delta_{\mathrm{XY}}^{XZ} = \left( \delta_{\mathrm{XY}}^{X} + \delta_{\mathrm{XY}}^{Z} \right) / 2$.
\paragraph{Pairing snapshots between bases.} 
Because the $X$- and $Z$-basis snapshots are independent experimental realizations, there is no unique pairing between individual snapshots acquired at the same sampling point.
We combine the two datasets by pairing snapshots within each sampling point according to their acquisition index, until the smaller dataset is exhausted (\fig~\ref{fig:number_of_snapshots}). This pairing is therefore arbitrary. In \app~\ref{app:xz_pairing_ablation}, we verify that the selected correlators and recovered decision boundary are robust to random re-pairings of the $X$- and $Z$-basis snapshots.

\section{Correlators}\label{app:correlators}
\subsection{Boolean function theory}\label{app:boolean_fourier}
The fact that snapshot data takes the form $S\in\{\pm1\}^{|\mathcal{I}|}$ where $\mathcal{I}$ is an index set, means that neural networks trained on this data are \textit{Boolean functions}. 
A basic result in Boolean function analysis is the Fourier expansion theorem, which states that any $f:\{\pm1\}^{|\mathcal I|}\rightarrow\mathbb{R}$ can be uniquely expressed as a multilinear polynomial:
\begin{equation}
     f( S ) =\sum_{A \subseteq \mathcal I}  c_A(f) \prod_{j \in A} S_j
\label{eq:multilinear_polynomial} 
\end{equation}
where the sum is over all {subsets} of $\mathcal{I}$, including the empty subset. $c_A$ are (real-valued) Fourier coefficients in the expansion, and $\prod_{j \in A} S_j$ are the basis functions (known in the literature as monomials, characters, or parity functions). 

As stated in the main text, we focus on convolutions of spatially arranged Boolean data. Following \eq~\eqref{eq:conv}, we define $f[P]_T = f[P](S)_T$ as a function depending on spins $\{S_i\}_{i\in T(P)}$ within a translated pattern $P$. With \eq~\eqref{eq:multilinear_polynomial} ($\mathcal I=P$, $A=P'$), this gives:
\begin{equation*}
\begin{aligned}
     f[P]_T &=\sum_{P' \subseteq P} c_{P'}(f) \prod_{i \in T(P')} S_{i}\\
     \end{aligned}
\end{equation*}
Applying a spatial average to this gives:
\begin{equation*}
\begin{aligned}
    \frac{1}{|\mathcal{T}_P|}\sum_T f[P]_T &=\frac{1}{|\mathcal{T}_P|}\sum_T \sum_{P' \subseteq P} c_{P'}(f)  \prod_{i \in T(P')} S_{i} \\
    &=\sum_{P'\subseteq P} c_{P'}(f)\,C[P'],
\end{aligned}     
\end{equation*}
which is the result used in the main text. Note that this requires $\mathcal{T}_{P'}=\mathcal{T}_{P}$ for all $P'$, i.e. correlators $C[P']$ need to be calculated over the same spins as $C[P]$ (e.g. $C[\square \bs]$ is only equivalent to $C[ \bs]$ up to edge effects).

\subsection{Counting patterns}\label{app:counting_correlators}
The number of patterns in 1D can be worked out as follows. It amounts to choosing $k$ positions from $\{ 1,...,D\}$ \textit{with translation invariance}; shifted patterns are equivalent. That is, instead of considering $k$ absolute sites, one can consider instead $k-1$ `spacings' (relative distances) between sites. Having fixed the first site (being equivalent to all others), choose $k-1$ from the $D-1$ remaining sites to define a pattern. The number of patterns of all orders on $D$ sites is thus:
$$
\sum_{k=1}^D \binom{D-1}{k-1} = 2^{D-1}
\label{eq:exponential_scaling}
$$
For 2D, finding the number of patterns is more complicated due to additional symmetries (rotations and reflections), but the number similarly scales exponentially and can be found using Burnside-Pólya Counting \cite{gross2008combinatorial}. The numbers in the main text were found instead by complete enumeration with a simple Python script (see \tab~\ref{tab:subpattern_numbers}).
\begin{table}[h]
\centering
\caption{Number of subpatterns for 3x3 and 4x4 filters and different symmetry groups.}
\begin{tabular}{lrr}
\toprule
\textbf{symmetries} & \textbf{3x3} & \textbf{4x4} \\
\midrule
-- & 511 & 65,535 \\
translations & 400 & 57,856 \\
rotations, reflections & 101 & 8,547 \\
translations, rotations, reflections & 85 & 7,625 \\
\bottomrule
\end{tabular}\label{tab:subpattern_numbers}
\end{table}

\section{Details of TetrisCNN architecture and hyperparameters}\label{app:archi_hyperparams}
\begin{figure*}[t]
    \centering
    \includegraphics[width=0.98\textwidth]{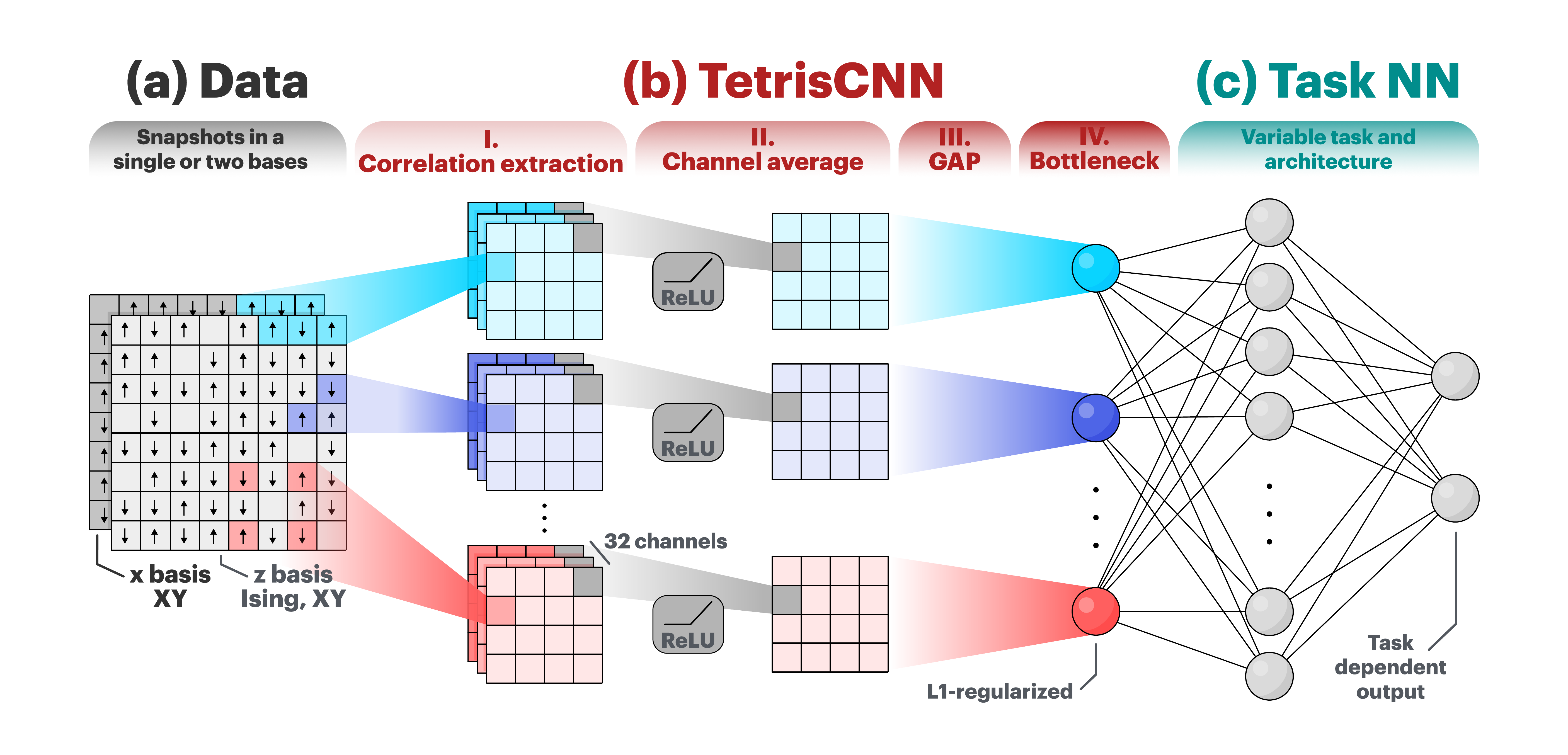}
    \caption{
    \figcap{Detailed architecture of TetrisCNN.} (a) A batch of $B$ input snapshots (or pairs of snapshots, when two measurement bases are combined) is processed by (b) $K$ parallel convolutional branches defined by patterns $P_k$. Each branch produces one scalar activation $z_k$ through two convolutional layers and global average pooling [\eq~\eqref{eq:branch}]. The first convolution sets the spatial pattern $P_k$ and is the only stage that reads correlations between sites; the $1\times1$ convolution acts site-by-site across channels and therefore cannot introduce any new spatial correlations. The resulting activations $\{z_k\}_{k=1}^K$ form the sparse interpretable bottleneck, on which the L1 penalty $\sum_k\lambda_k|z_k|$ acts. (c) The bottleneck is then read out by a task network (here, a small fully connected network) whose architecture is adapted to the learning task.}
\label{fig:TetrisCNN_archi}
\end{figure*}

This Appendix complements the architectural description of TetrisCNN in
\seclab~\ref{ss:interpretable_Tetris}. A detailed schematic of the architecture
is shown in \fig~\ref{fig:TetrisCNN_archi}, while the hyperparameters used for
the classifiers in the main text are summarized in
\tab~\ref{tab:hyperparams}.

\begin{table}[t]
\centering
\caption{\figcap{Hyperparameters of the two classification models reported in the main text.} The columns correspond to the Ising and XY classifiers shown in panels (b) and (f) of \fig~\ref{fig:interpretable_detection}, respectively. Entries spanning both columns are shared between the two models.}
\label{tab:hyperparams}
\resizebox{\linewidth}{!}{%
\begin{tabular}{lll} 
\toprule
\textbf{Dataset} & \textbf{TFIM ($Z$ basis)} & \textbf{XY ($Z,X$ bases)} \\ 
\midrule
\textbf{TetrisCNN} & \multicolumn{2}{c}{} \\
Available branches & \multicolumn{2}{c}{
\begin{tabular}[c]{@{}r@{}}\smallskip
\([\,\blacksquare, \text{stride=1},\text{dilation=1}\,]\)\quad\quad\quad\,\,~~\\\smallskip
\([\,\blacksquare\blacksquare, \text{stride=1},\text{dilation=1},\text{rot=C4}\,]\)~\\\smallskip
\(\qty[\,\substack{\blacksquare\square\\\square\blacksquare}, \text{stride=1},\text{dilation=1},\text{rot=C4}\,]\)~\\\smallskip
\(\qty[\,\substack{\blacksquare\blacksquare\\\blacksquare\square}
, \text{stride=1},\text{dilation=1},\text{rot=C4}\,]\)~\\\smallskip
\(\qty[\,\substack{\blacksquare\blacksquare\\\blacksquare\blacksquare}, \text{stride=1},\text{dilation=1},\text{rot=C4}\,]\)
\end{tabular}} \\
Bottleneck size (\# branches) & \multicolumn{2}{c}{5} \\
$\lambda_{\min}$, $\lambda_{\max}$ & \multicolumn{2}{c}{$10^{-3}$, $10^3$} \\ 
\midrule
\textbf{Task NN} & \multicolumn{2}{c}{} \\
Task type & \multicolumn{2}{c}{Classification} \\
Architecture & \multicolumn{2}{c}{Multilayer Perceptron} \\
Layers & \multicolumn{2}{c}{{[}\# branches, 32, 16, 2]} \\ 
\midrule
\textbf{Training parameters} & \multicolumn{2}{c}{} \\
Loss function & \multicolumn{2}{c}{Cross Entropy Loss} \\
Optimizer & \multicolumn{2}{c}{AdamW} \\
Weight decay & \multicolumn{2}{c}{1e-05} \\
Learning rate scheduler & \multicolumn{2}{c}{Reduce on Plateau} \\
\hspace{0.5cm}Initial learning rate & \multicolumn{2}{c}{1e-2} \\
\hspace{0.5cm}Reduce factor & \multicolumn{2}{c}{0.5} \\
\hspace{0.5cm}Patience & \multicolumn{2}{c}{5} \\
\hspace{0.5cm}Minimal learning rate & \multicolumn{2}{c}{1e-9} \\
Max epochs & \multicolumn{2}{c}{250} \\
Early Stopping & \multicolumn{2}{c}{Yes} \\
\hspace{0.5cm}Warmup & \multicolumn{2}{c}{150} \\
\hspace{0.5cm}Patience & \multicolumn{2}{c}{10} \\
RNG seed & \multicolumn{2}{c}{ 42} \\
Batch size & \multicolumn{2}{c}{64} \\
\bottomrule
\end{tabular}}
\end{table}

\paragraph{Task network}
The task neural network is a flexible readout of the interpretable bottleneck and is not intrinsic to TetrisCNN. For the classification and regression tasks considered here, we use a multilayer perceptron. We tested architectures with one to three hidden layers and found no dependence of performance on the task network architecture. We therefore use a two-hidden-layer fully-connected network with architecture $[\,\#\text{branches},32,16,2\,]$ for classification.

\paragraph{Branch penalty}
As defined in \eq~\eqref{eq:branch_penalty}, the branch penalty $\lambda_k$ increases with the pattern cardinality $|P_k|$ through a log-linear interpolation between $\lambda_{\min}$ and $\lambda_{\max}$, favoring lower-order correlators. More generally, the branch penalty provides a way to encode what is considered a \textit{simple} physical description. For example, when including dilated patterns, the penalty can increase with spatial extent to favor local correlators. Similarly, non-equivariant branches can be penalized more strongly than their symmetry-constrained counterparts to favor symmetric descriptions. Thus, the penalty can be adapted to encode prior knowledge about the complexity of candidate physical observables.

\section{Separation of active and deactivated branches by learning-rate annealing}\label{app:bottleneck_optimization}
\subsection{Origin of the activation floor}
\label{app:noise_floor}

Everything we read off the TetrisCNN bottleneck relies on distinguishing active branches from deactivated ones. In practice, deactivated branches do not decay exactly to zero but settle onto a small residual floor, visible as the nearly flat lines at the bottom of \fig~\ref{fig:interpretable_detection}(b,e). Here we explain the origin of this floor and why learning-rate annealing is required to clearly separate active and deactivated branches, yielding a sparse bottleneck.

Let $\mathbf{z}\in\mathbb{R}^K$ denote the bottleneck activations of the $K$ parallel branches for a given input [cf.~\eq~\eqref{eq:branch}]. The network is trained with the task loss together with the branch-wise sparsity penalty,
\begin{equation}
    \mathcal{L}
    =
    \mathcal{L}_{\mathrm{Task}}(\hat y,y)
    +
    \sum_{k=1}^{K}\lambda_k |z_k|\,,
    \label{eq:total_loss}
\end{equation}
where $\lambda_k>0$ is the penalty on branch $k$. To expose the mechanism behind the activation floor, we first treat the bottleneck activations $z_k$ as effective optimization variables. In the actual network, AdamW updates the branch parameters rather than $z_k$ directly, but parameter updates induce corresponding changes in $z_k$ through the branch Jacobian. We return to this point later. Therefore, differentiating with respect to a single bottleneck activation gives
\begin{equation}
    \nabla_{z_k}\mathcal{L}
    =
    \underbrace{\nabla_{z_k}\mathcal{L}_{\mathrm{Task}}}_{\text{task signal}}
    +
    \underbrace{\lambda_k\,\mathrm{sgn}(z_k)}_{\text{sparsity term}}\,.
    \label{eq:gradient_split}
\end{equation}
The second term always pushes the activation toward zero.

Once the typical task contribution of a branch becomes negligible compared with
its sparsity penalty,
\begin{equation}
    \operatorname{Avg}_{\mathcal{B}}
    \left[
    \left|
    \nabla_{z_k}\mathcal{L}_{\mathrm{Task}}
    \right|
    \right]
    \ll \lambda_k,
    \label{eq:floor_cond}
\end{equation}
the gradient entering the optimizer is dominated by the $L_1$ term. For an
optimization step $t$ using minibatch $\mathcal{B}_t$, we can therefore write
\begin{equation}
    g_t
    =
    \nabla_{z_k}\mathcal{L}^{(t)}
    \approx
    \lambda_k\,\mathrm{sgn}(z_k^{(t)}).
    \label{eq:penalty_gradient}
\end{equation}

For ordinary gradient descent, this would give
\begin{equation}
    \Delta z_k^{(t)}
    =
    -\alpha_t g_t
    \approx
    -\alpha_t\lambda_k\,\mathrm{sgn}(z_k^{(t)}),
\end{equation}
so the magnitude of the update would scale directly with $\lambda_k$ and instantaneous learning rate $\alpha_t$. Adam or AdamW (which differs from Adam only in how weight decay is applied)~\cite{kingma_adam_2017,loshchilov2017decoupled} behave differently because they normalize the gradient by the square root of its running second moment. Ignoring, for the moment, the first-moment dynamics, the Adam update has the schematic form
\begin{equation}
    \Delta z_k^{(t)}
    \sim
    -\alpha_t
    \frac{g_t}{\sqrt{v_t}},
    \label{eq:adam_schematic}
\end{equation}
where $v_t$ is the exponential moving average of $g_t^2$.

For the penalty-dominated gradient of \eq~\eqref{eq:penalty_gradient}, its magnitude is approximately constant,
\begin{equation}
    |g_t| \approx \lambda_k,
    \qquad
    g_t^2 \approx \lambda_k^2.
\end{equation}
If this penalty-dominated regime persists sufficiently long, Adam's running second-moment estimate approaches the squared gradient scale,
\begin{equation}
    v_t \sim \lambda_k^2,
    \qquad
    \sqrt{v_t}\sim \lambda_k.
\end{equation}
Substituting this into \eq~\eqref{eq:adam_schematic} gives
\begin{equation}
    \Delta z_k^{(t)}
    \sim
    -\alpha_t
    \frac{\lambda_k\,\mathrm{sgn}(z_k^{(t)})}
         {\lambda_k}
    =
    -\alpha_t\,\mathrm{sgn}(z_k^{(t)}).
    \label{eq:noise_floor_step}
\end{equation}

The important point is that the leading dependence on $\lambda_k$ cancels. The sparsity strength still determines whether the branch enters this penalty-dominated regime, but once it does, increasing $\lambda_k$ no longer substantially reduces the residual update scale under Adam.

Near zero, these updates repeatedly change the sign of $z_k$. For example, if $0<z_k<\alpha_t$, the next update can carry it across zero. The sign of the $L_1$ gradient then reverses and the following updates push it back in the opposite direction. The activation therefore fluctuates around zero instead of converging smoothly to it. The characteristic scale of these residual fluctuations is set by the instantaneous learning rate,
\begin{equation}
    z_{\mathrm{floor}}
    =
    \mathcal{O}(\alpha_t).
    \label{eq:noise_floor}
\end{equation}

In the full network, AdamW updates the branch parameters $\boldsymbol{\theta}_k$ rather than the activation $z_k$ directly. In the penalty-dominated regime, the gradient of the sparsity term with respect to a branch parameter $\theta_{k,j}$ is, by the chain rule,
\begin{equation}
    \frac{\partial \mathcal{L}_{L_1}}{\partial \theta_{k,j}}
    =
    \lambda_k\,\mathrm{sgn}(z_k)
    \frac{\partial z_k}{\partial \theta_{k,j}} .
\end{equation}
The corresponding Adam second-moment estimate therefore scales as $\lambda_k^2(\partial z_k/\partial\theta_{k,j})^2$. As in the simplified activation-level argument, the overall factor $\lambda_k$ cancels between the gradient and its normalization. The resulting parameter update is therefore of order $\alpha_t$, with a prefactor determined by the local derivatives of $z_k$ with respect to the branch parameters. For a sufficiently small parameter update,
\begin{equation}
    \Delta z_k
    \approx
    \nabla_{\boldsymbol{\theta}_k} z_k
    \cdot
    \Delta\boldsymbol{\theta}_k
    =
    \mathcal{O}(\alpha_t).
\end{equation}
Thus, moving from the simplified activation-level picture to the actual parameter updates changes the precise prefactor but preserves the central result: once the sparsity term dominates, the residual activation scale is controlled by the learning rate rather than the overall magnitude of
$\lambda_k$.

\subsection{Active versus deactivated branches}
\label{app:survival_threshold}

A branch remains active when its contribution to the task is sufficiently large to counteract the sparsity penalty. Near convergence, this requires the task and sparsity contributions to approximately balance,
\begin{equation}
    \operatorname{Avg}_{\mathcal{B}}
    \left[
    \left|
    \nabla_{z_k}\mathcal{L}_{\mathrm{Task}}
    \right|
    \right]
    \sim
    \lambda_k .
    \label{eq:balance}
\end{equation}
If instead the task signal becomes much smaller than $\lambda_k$, the branch is driven toward the learning-rate-dependent floor described in \seclab~\ref{app:noise_floor}.

This is the mechanism by which the branch penalties perform model selection. Increasing $\lambda_{\max}$ raises the penalty on the more complex patterns, so branches whose contribution to the task is insufficient are progressively suppressed. In practice, the surviving set rapidly becomes stable over a broad range of $\lambda_{\max}$ while the classification accuracy remains essentially unchanged, as shown in \fig~\ref{fig:lambdamax} of \app~\ref{app:robustness}. We therefore identify a branch as active when its converged activation remains well above the residual floor set by the final learning rate.

\subsection{Sharpening sparsity with learning-rate annealing}
\label{app:lr_scheduling}

\begin{figure}[htbp]
    \centering
    \includegraphics[width=0.99\linewidth]{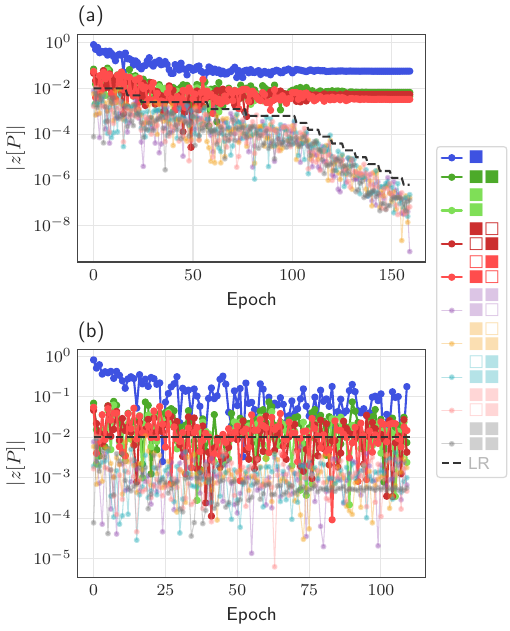}
    \caption{\figcap{Learning-rate annealing separates active and deactivated branches.} Training history of the XY classifier [rotation unconstrained variant of panel (e) of \fig~\ref{fig:interpretable_detection}] with all settings kept fixed except for the learning-rate schedule. (a) With ReduceLROnPlateau, the deactivated branches follow the learning-rate scale and decrease whenever it is reduced, consistent with the $\mathcal{O}(\alpha_t)$ scaling of \eq~\eqref{eq:noise_floor}. (b) With the learning rate held fixed at $\alpha = 10^{-2}$, the residual floor remains at the same order of magnitude and overlaps with the weakest active branches.}
    \label{fig:noise_floor}
\end{figure}

Because the activation floor scales as $\mathcal{O}(\alpha_t)$, annealing the learning rate progressively separates deactivated branches from active ones. This is shown directly in \fig~\ref{fig:noise_floor}(a) for the XY classifier. The suppressed branches track the learning-rate scale and decrease whenever the learning rate is reduced, consistent with the scaling of \eq~\eqref{eq:noise_floor_step}.

When the learning-rate schedule is disabled, \fig~\ref{fig:noise_floor}(b), the learning rate remains fixed at $\alpha = 10^{-2}$, and the branch activation floor correspondingly stays at the same $\mathcal{O}(10^{-2})$ scale, overlapping with the weakest active branches. The sparse bottleneck then becomes much harder to identify. Learning-rate annealing is therefore essential in practice for obtaining a clear separation between active and deactivated branches.

We compared cosine annealing, exponential decay, one-cycle scheduling, and ReduceLROnPlateau using the same training budget. All four produced a clear separation between active and deactivated branches by the end of training, suggesting that the key ingredient is annealing itself rather than the schedule's precise shape. Throughout this work, we use ReduceLROnPlateau, halving the learning rate whenever the validation metric does not improve for five epochs.

\section{Unsupervised detection of transitions in the experimental sweeps: method comparison}\label{app:unsupervised_comparison}
To define the supervised classification labels used by TetrisCNN, we first estimate the transition location in each experimental dataset using five established unsupervised approaches: learning by confusion (LBC)~\cite{van_nieuwenburg_learning_2017}, the prediction-divergence method (PDM)~\cite{schafer_vector_2019,greplova_unsupervised_2020,arnold_interpretable_2021}, principal component analysis (PCA)~\cite{Wang16}, Uniform Manifold Approximation and Projection (UMAP)~\cite{mcinnes_umap_2020}, and distance learning (DL)~\cite{malyshev2026distance-fb9}. For consistency, all transition estimates are reported as the interval between the two acquisition times flanking the detected change. For LBC, this interval is additionally enlarged when the inferred transition location varies across random initializations or sparsity strengths. The resulting estimates are summarized in \fig~\ref{fig:interpretable_detection}(a,d).

\subsection{Learning by confusion}
\label{app:lbc}

Learning by confusion (LBC)~\cite{van_nieuwenburg_learning_2017} locates a transition by repeatedly imposing artificial binary labels on the data. For a trial transition point $y'$, snapshots are labeled according to whether their parameter lies below or above $y'$, and a classifier is trained to distinguish the two groups. Sweeping $y'$ across the experimental range gives the characteristic confusion curve, whose interior accuracy maximum estimates the transition,
\begin{equation}
    y_c \approx \arg\max_{y'} \mathrm{Acc}(y').
    \label{eq:lbc}
\end{equation}
We use TetrisCNN itself as the classifier for each trial split.

\begin{figure}[t]
    \centering
    \includegraphics[width=\linewidth]{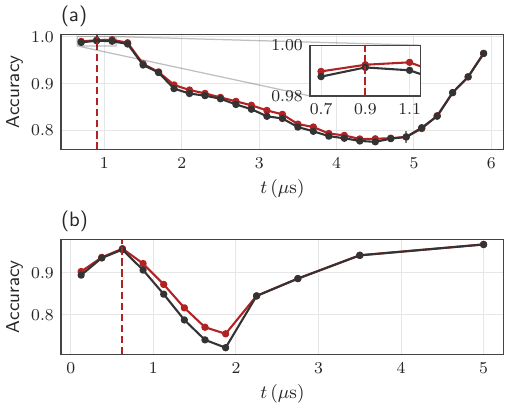}
    \caption{\figcap{Learning-by-confusion transition estimate.}
    Training (red) and validation (black) classification accuracy as a function of the hypothesized split time for (a) the Ising and (b) the XY datasets. The red dashed line marks the interior accuracy maximum of \eq~\eqref{eq:lbc}. Accuracies are averaged over five random initializations; most error bars are smaller than the markers. For XY, the maximum gives $t\approx0.625\us$, bracketed by $[0.5,0.75]\us$. For Ising, the maximum is broad and shifts between neighboring partitions, yielding $t\approx0.9\us$ within $[0.8,1.2]\us$. Because the experimental sweeps start close to the central confusion maximum, only the second half of the characteristic ``W'' curve is resolved.}
    \label{fig:benchmark_lbc}
\end{figure}

\Fig~\ref{fig:benchmark_lbc} shows the resulting training and validation accuracies. For the XY dataset, the interior maximum is well defined and consistently falls on the third acquisition-time partition, corresponding to $t\approx0.625\us$ and bracketed by $[0.5,0.75]\us$. For the Ising dataset, the maximum is broad and shallow: its location shifts between neighboring partitions depending on the random initialization, sparsity strength $\lambda_{\max}$, and whether training or validation accuracy is used. Across eleven values of $\lambda_{\max}$ and five random initializations per value, we therefore report the median maximum, $t\approx0.9\us$, and use the envelope of the inferred transition locations to obtain the broader interval $[0.8,1.2]\us$.

\subsection{Prediction-divergence method}
\label{app:pdm}

The prediction-divergence method (PDM)~\cite{schafer_vector_2019,greplova_unsupervised_2020,arnold_interpretable_2021} trains a single regressor to predict an experimental tuning parameter $y$ from a snapshot $x$,
\begin{equation}
    \hat y=f_{\theta}(x).
\end{equation}
The transition is then identified from the response of the average prediction to the true tuning parameter,
\begin{equation}
    I(y)
    =
    \frac{\partial \langle \hat y\rangle_y}{\partial y},
    \label{eq:phase_indicator}
\end{equation}
where $\langle \hat y\rangle_y$ denotes the average network prediction over all snapshots acquired at the same true parameter value $y$. The maximum of $I(y)$ marks the point at which the learned prediction changes most rapidly with $y$. For multivariate tuning parameters, the corresponding quantity is the divergence~\cite{schafer_vector_2019}. We use TetrisCNN as the regressor and train it with mean-squared error, with regression quality evaluated using the aggregated $R^2$ defined in \app~\ref{app:multivariate_R2}.

\begin{figure}[t]
    \centering
    \includegraphics[width=\linewidth]{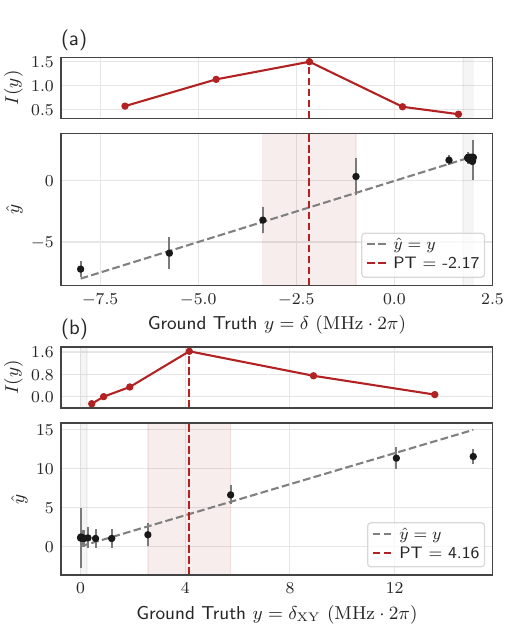}
    \caption{\figcap{Prediction-divergence transition estimate.}
    Phase indicator $I(y)$ (top) and average prediction $\langle\hat y\rangle_y$ against the true tuning parameter (bottom) for (a) the Ising dataset, regressing $\longitudinal$, and (b) the XY dataset, regressing $\deltaxy$. The red dashed line marks the maximum of the first-order finite-difference indicator. The corresponding transition estimates are $t\approx1.1\us$, bracketed by $[1.0,1.2]\us$ (Ising), and $t\approx0.625\us$, bracketed by $[0.5,0.75]\us$ (XY). In experimental parameter space, the maxima occur at $\longitudinal/2\pi\approx-2.2\,\mathrm{MHz}$ and $\deltaxy/2\pi\approx4.2\,\mathrm{MHz}$, respectively.}
    \label{fig:benchmark_pdm}
\end{figure}

As shown in \fig~\ref{fig:benchmark_pdm}, the indicator peaks at the third interval, corresponding to $t\approx1.1\us$ for the Ising dataset and to $t\approx0.625\us$ for the XY dataset. For Ising we use regression of $\longitudinal$; for XY we regress $\deltaxy$. We evaluate \eq~\eqref{eq:phase_indicator} using a first-order finite difference, so the maximum naturally lies between two neighboring acquisition times. These two times define the uncertainty interval, giving $[1.0,1.2]\us$ for Ising and $[0.5,0.75]\us$ for XY.

\subsection{Dimensionality reduction and clustering}
\label{app:dimred}

\begin{figure*}[t]
    \centering
    \includegraphics[width=\textwidth]{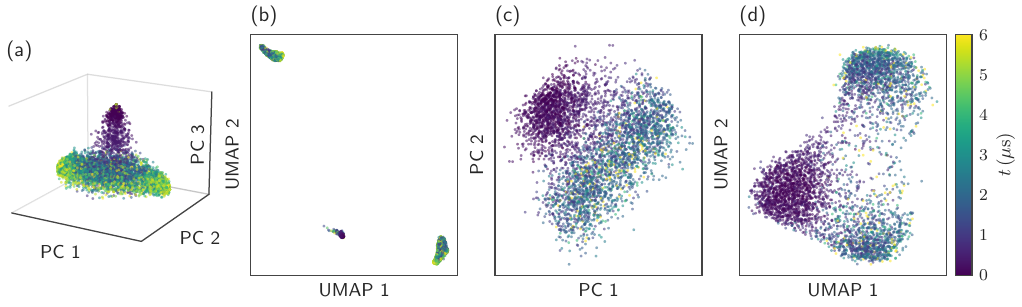}
    \caption{\figcap{Low-dimensional embeddings of the experimental snapshots.}
    (a,b) Ising and (c,d) XY snapshots projected with PCA (a,c) and UMAP (b,d), colored by acquisition time. In all cases, the earliest acquisition times form a distinct group that is isolated by $k$-means clustering. The later-time snapshots overlap substantially and their cluster assignments no longer follow the sweep monotonically.}
    \label{fig:benchmark_clustering}
\end{figure*}

As a complementary approach, we project the raw experimental snapshots into a low-dimensional representation and cluster the resulting embeddings. We use PCA~\cite{Wang16}, which projects the data onto directions of maximal variance, and UMAP~\cite{mcinnes_umap_2020}, which constructs a nonlinear embedding that approximately preserves local neighborhood structure. The projected snapshots are clustered with $k$-means, and each acquisition time is assigned the majority cluster of its snapshots. We identify the transition where this majority assignment first changes.

\Fig~\ref{fig:benchmark_clustering} shows the resulting embeddings. For both datasets, PCA and UMAP isolate the earliest acquisition times from the remainder of the sweep. In the Ising data, the initial cluster contains the majority of snapshots up to $t=1.0\us$ but not at $1.2\us$, giving the interval $[1.0,1.2]\us$. In the XY data, the corresponding change occurs between $0.5\us$ and $0.75\us$. At later times, the cluster assignments no longer follow the sweep monotonically, so we use the embeddings only to identify this primary change.

UMAP resolves the early-time cluster in two dimensions for both datasets. Two principal components are also sufficient for XY, whereas Ising requires three principal components before the early-time cluster is separated cleanly. The corresponding silhouette scores and clustering settings are listed in \tab~\ref{tab:clustering}.

\begin{table}[t]
    \centering
    \caption{\figcap{Clustering of low-dimensional snapshot embeddings.}
    For each dataset and dimensionality-reduction method, we report the smallest embedding and number of $k$-means clusters for which the initial-time cluster is isolated, together with the silhouette score and the transition interval inferred from the majority-cluster assignment. UMAP is run with $\texttt{n\_neighbors}=250$, $\texttt{min\_dist}=0$, and the cosine metric.}
    \label{tab:clustering}
    \resizebox{\columnwidth}{!}{%
    \begin{tabular}{llccc}
        \toprule
        \textbf{Dataset} & \textbf{Reducer} & \textbf{Components}, $k$ & \textbf{Silhouette} & \textbf{Transition interval} \\
        \midrule
        Ising & UMAP & $2$, $k=3$ & $0.93$ & $1.0$--$1.2\us$ \\
        Ising & PCA  & $3$, $k=4$ & $0.48$ & $1.0$--$1.2\us$ \\
        XY ($X,Z$) & UMAP & $2$, $k=3$ & $0.62$ & $0.5$--$0.75\us$ \\
        XY ($X,Z$) & PCA  & $2$, $k=3$ & $0.51$ & $0.5$--$0.75\us$ \\
        \bottomrule
    \end{tabular}%
    }
\end{table}

\subsection{Distance learning}
\label{app:dl}

The distance-learning (DL) method of Malyshev \textit{et al.}~\cite{malyshev2026distance-fb9} compares the full snapshot distributions at different parameter values rather than embedding individual snapshots. A discriminator network is trained to estimate pairwise squared Hellinger $f$-divergences between the empirical snapshot distributions. The resulting divergence matrix is then treated as a distance matrix and clustered with HDBSCAN~\cite{campello2013density-based}; a transition is identified at the boundary between the resulting clusters.

We apply the implementation accompanying \reflab~\cite{malyshev2026distance-fb9} to the experimental datasets. In contrast to the dense parameter grids used in the original numerical benchmarks, the experimental sweeps contain only a small number of discrete acquisition times. We therefore set the HDBSCAN minimum cluster size to $2$, the smallest value that still yields meaningful clusters in this setting.

\begin{figure*}[htbp]
    \centering
    \includegraphics[width=\textwidth]{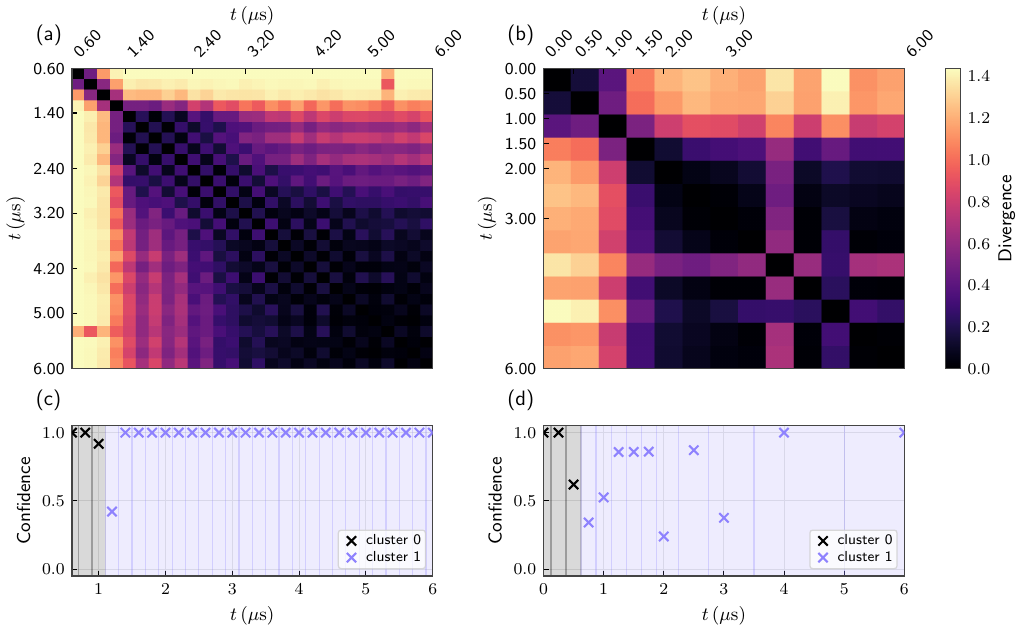}
    \caption{\figcap{Distance-learning transition estimate.}
    (a,b) Pairwise squared Hellinger $f$-divergence matrices between the snapshot distributions at different acquisition times and (c,d) the corresponding HDBSCAN cluster assignments and confidence for the Ising (a,c) and XY (b,d) datasets. The earliest acquisition points form a distinct cluster, yielding transition estimates of $t\approx1.1\us$ within $[1.0,1.2]\us$ for Ising and $t\approx0.625\us$ within $[0.5,0.75]\us$ for XY. Figure generated using code adapted from the repository accompanying \reflab~\cite{malyshev2026distance-fb9}.}
    \label{fig:benchmark_dl}
\end{figure*}

As shown in \fig~\ref{fig:benchmark_dl}, DL isolates the earliest acquisition points from the remainder of each sweep. The resulting boundary lies at $t\approx1.1\us$ for Ising and $t\approx0.625\us$ for XY. As for the dimensionality-reduction methods, the uncertainty is determined by the discrete acquisition grid and spans the two times flanking the cluster boundary: $[1.0,1.2]\us$ for Ising and $[0.5,0.75]\us$ for XY.

\subsection{Comparison of transition estimates}
\label{app:comparison}

All five estimates identify the same primary change in each dataset. For the Ising sweep, PDM, PCA, UMAP, and DL place the transition at $t\approx1.1\us$, bracketed by $[1.0,1.2]\us$. LBC gives a compatible but broader estimate centered at $t\approx0.9\us$ with an interval $[0.8,1.2]\us$, reflecting the flat confusion maximum and its variation across fits. For the XY sweep, all five methods agree on $t\approx0.625\us$, bracketed by $[0.5,0.75]\us$.

This agreement defines the locations used to construct the supervised classification labels in the main TetrisCNN analysis. For the Ising dataset, the detected change corresponds to the loss of the initially strong $Z$-polarization rather than to the later emergence of antiferromagnetic order, as discussed in \seclab~\ref{sec:discussion}. For the XY dataset, all methods consistently identify the same change near $t\approx0.625\us$.

\section{Supporting material}\label{app:supporting_material}
\begin{figure*}[ht]
    \centering
    \includegraphics[width=0.98\textwidth]{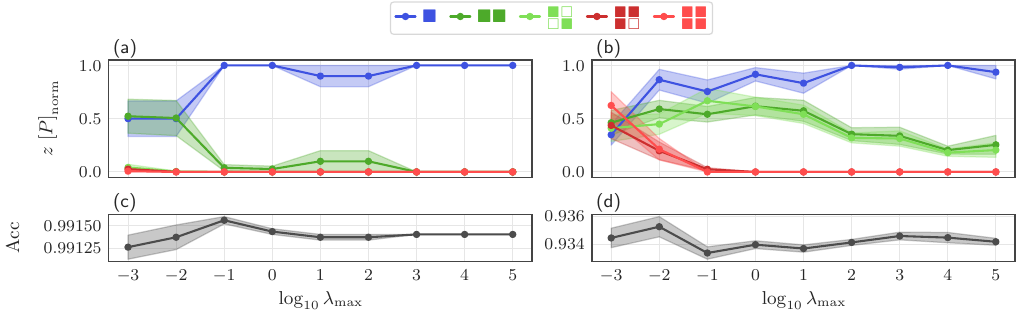}
    \caption{\figcap{Robustness of branch selection to the sparsity regularization strength $\lambda_{\max}$.} (a,b) Normalized bottleneck activations and (c,d) validation classification accuracy as a function of the maximum sparsity coefficient $\lambda_{\max}$ for the Ising (a,c) and XY (b,d) datasets. Results are averaged over ten random initializations.}
    \label{fig:lambdamax}
\end{figure*}

\begin{figure*}[ht]
    \centering
    \includegraphics[width=0.98\textwidth]{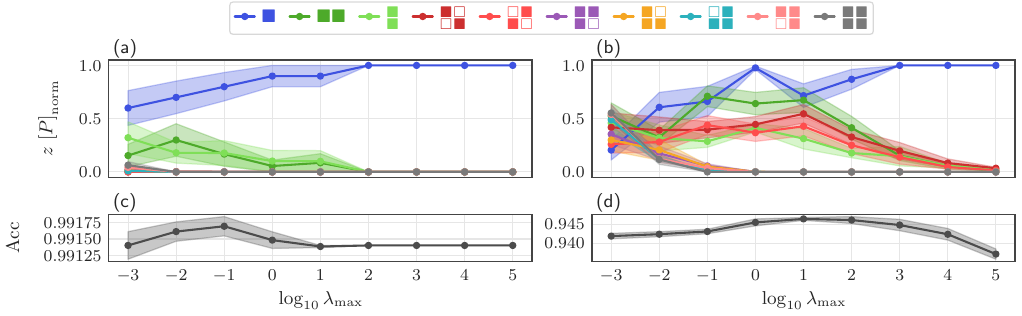}
    \caption{\figcap{Robustness of branch selection to the sparsity regularization strength $\lambda_{\max}$ for filters that are not rotationally invariant.} (a,b) Normalized bottleneck activations and (c,d) validation classification accuracy as a function of the maximum sparsity coefficient $\lambda_{\max}$ for the Ising (a,c) and XY (b,d) datasets. Results are averaged over ten random initializations.}
    \label{fig:lambdamax_notC4}
\end{figure*}

\subsection{Robustness of the learned descriptors to regularization}\label{app:robustness}

A key component of TetrisCNN is the L1 regularization applied to the bottleneck activations,
\begin{equation}
\loss_{\mathrm{L1}}=\sum_k \lambda_k |\actbottle_k|,
\end{equation}
which encourages the network to identify a sparse set of descriptors sufficient for solving the classification task. Specifically, $\lambda_k$ interpolates log-linearly between $\lambda_{\min}$ and $\lambda_{\max}$ according to the pattern cardinality $|P_k|$ [\eq~\eqref{eq:branch_penalty}]. To investigate the effect of the regularization strength, we vary the maximum penalty $\lambda_{\max}$ while keeping $\lambda_{\min} = 10^{-3}$ and the interpolation fixed. This hyperparameter controls the simplicity of the learned representation: increasing $\lambda_{\max}$ penalizes larger-filter branches more strongly and therefore favors solutions based on fewer and lower-order correlators. \Fig~\ref{fig:lambdamax} summarizes the resulting branch activations and classification accuracies, averaged over ten random initializations.

For the Ising dataset, \fig~\ref{fig:lambdamax}(a), the same $1\times1$ branch is consistently selected across a broad range of $\lambda_{\max}$, once the regularization exceeds a threshold around $\lambda_{\max}\approx10^{-1}$. Then the network largely suppresses all remaining branches and converges to the one-dimensional representation discussed in the main text. Importantly, this simplification has almost no effect on predictive performance: the classification accuracy remains essentially constant throughout the entire range of sparsity strengths considered, \fig~\ref{fig:lambdamax}(c). The value $\lambda_{\max}=10^3$ used in the figures in the main text lies within this stable regime.

For the XY dataset, we see a similar behavior. After reaching the penalty threshold at $\lambda_{\max}\approx 10^{-1}$, the network consistently suppresses the redundant branches while retaining the same three dominant descriptors identified in the main text. Once again, this reduction in representation complexity causes only a negligible change in classification accuracy, as seen in \fig~\ref{fig:lambdamax}(d), and the main-text value of $\lambda_{\max}=10^3$ lies within this stable region.

Taken together, these results demonstrate that the physically interpretable descriptors identified by TetrisCNN are not sensitive to the precise choice of sparsity strength. Instead, the sparsity regularization acts primarily as a model-selection mechanism, selecting a simple representative from a large family of functionally equivalent solutions with nearly identical predictive performance. This observation is consistent with the Rashomon effect \cite{pmlr-v235-rudin24a}: many distinct internal representations achieve comparable classification accuracy, while the sparsity penalty biases the network toward the simplest among them.

\begin{figure*}[t]
    \centering
    \includegraphics[width=0.98\textwidth]{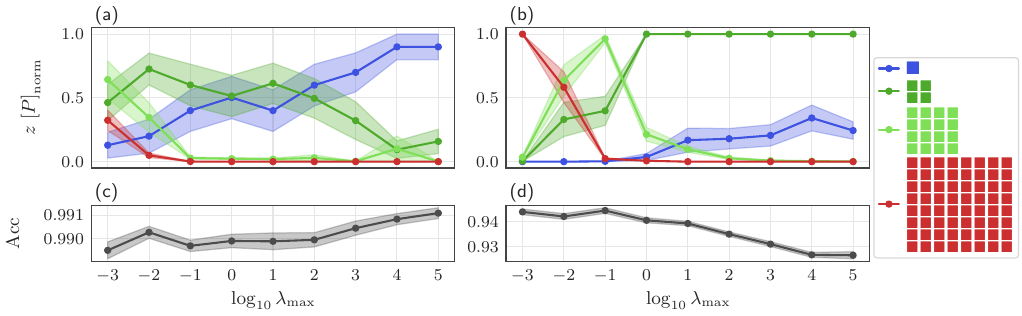}
    \caption{\figcap{Searching for the relevant scale} for the Ising (first column) and XY model (second column). (a,b) Normalized bottleneck activations and (c,d) validation classification accuracy as a function of the maximum sparsity coefficient $\lambda_{\max}$ for the Ising (a,c) and XY (b,d) datasets. Averaged over 10 random initializations.
    }\label{fig:big_kernels}
\end{figure*}

\subsection{Searching for the relevant scale}\label{app:search_for_scale}

The main text restricts the branch dictionary to patterns no larger than $2\times2$ (\tab~\ref{tab:Tetris_correlators}). Before committing to this choice, we tested whether TetrisCNN would favor a larger spatial scale if given the option. We repeat the same classification setup and $\lambda_{\max}$ sweep used to fix the label boundary (\app~\ref{app:lbc}), but replace the branch dictionary with four deliberately coarse-grained kernels: $1\times1$, $2\times2$, $4\times4$, and the full system size ($8\times8$ for the Ising data, $6\times7$ for the XY data). \Fig~\ref{fig:big_kernels} shows the resulting branch activations and validation accuracy as a function of $\lambda_{\max}$, averaged over ten random initializations. For both datasets, the network consistently suppresses kernels larger than $2\times2$ once the sparsity penalty exceeds a modest threshold with little effect on classification accuracy. For the XY data, suppressing the $4\times 4$ branch is accompanied by a small decrease in classification accuracy of approximately 1\%, suggesting this branch captures some useful nonlocal information. Nevertheless, the vast majority of the predictive signal comes from the local branches. These results indicate that local patterns are sufficient to capture nearly all information relevant to the classification tasks considered here, justifying our restriction to patterns no larger than $2 \times 2$ in the main text.

\subsection{Rotationally invariant and unconstrained TetrisCNN}\label{app:rotational_invariance}

To test the effect of imposing the $C_4$ rotational symmetry of the square lattice, we compare the rotationally invariant TetrisCNN used in the main text with an unconstrained variant across the same sweep of $\lambda_{\max}$. As shown in \fig~\ref{fig:lambdamax_notC4}, the unconstrained variant identifies the same types of relevant branches as the rotationally invariant network (\fig~\ref{fig:lambdamax}). For Ising, the learned representations are identical. For XY, the rotationally invariant model retains three branches and the unconstrained model retains five, corresponding to separate rotational variants of the same patterns. We therefore focus on the rotationally invariant network in the main text, as it provides an equivalent but lower-dimensional and simpler representation.

\begin{figure}[b]
    \centering
    \includegraphics[width=0.98\columnwidth]{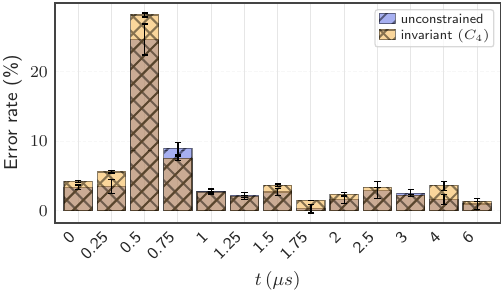}
    \caption{\figcap{Classification validation error across the XY sweep for rotationally invariant and unconstrained TetrisCNN.} The unconstrained network achieves approximately $1\%$ lower validation error, with the improvement concentrated near the phase transition around $0.625\,\mu$s.
    } \label{fig:rot_vs_no_rot_errors}
\end{figure}

Interestingly, the best unconstrained network trained on the XY dataset in \fig~\ref{fig:lambdamax_notC4}(b),(d) achieves approximately one percentage point higher validation accuracy than the best rotationally invariant network. As shown in \fig~\ref{fig:rot_vs_no_rot_errors}, this improvement is concentrated near the phase transition, indicating that the experimental snapshots contain some additional classification-relevant information that is not rotationally invariant. Nevertheless, the small accuracy difference and agreement in the selected pattern types support our use of the rotationally invariant model for the physical interpretation.

\subsection{Robustness to random pairings of \texorpdfstring{$X$}{X} and \texorpdfstring{$Z$}{Z} snapshots}
\label{app:xz_pairing_ablation}

The $X$- and $Z$-basis measurements in the XY dataset correspond to independent experimental realizations. Consequently, pairing individual $X$- and $Z$-basis snapshots into two-channel inputs is arbitrary. To verify that the interpretation reported in the main text does not depend on this choice, we repeat the complete TetrisCNN analysis for ten independent random re-pairings of the two measurement bases, using the same hyperparameters and sparsity strength as in the main-text analysis. For each re-pairing, we train TetrisCNN from scratch, identify the surviving branches and their leading Boolean-Fourier components, and independently fit the quadratic decision boundary in the resulting bottleneck space, as described in \app~\ref{app:dec_bound_log_reg_fit}. Because multiplying a decision-boundary equation by a nonzero constant leaves the boundary unchanged, we normalize the fitted equations to the coefficient of $C_{\mathrm{rot}}^Z[\bs\bs]$ before comparing their coefficients. We then report the mean and standard deviation of each normalized coefficient across the ten re-pairings in \tab~\ref{tab:XZ_random_repairings}, compare them with the main-text fit in \eq~\eqref{eq:decision_boundary_in_correlators}, and quantify the deviation of each main-text coefficient from the re-pairing mean in units of the corresponding standard deviation.

\begin{table}[b]
    \centering
    \begin{tabular}{c|c|c|c|c|c}
       Terms  & $C^X[\bs]^2$ & $C^X[\bs]$ & $C_{\mathrm{rot}}^Z[\bs\bs]$ & $C_{\mathrm{rot}}^Z[\sus{\bs\square\\\square\bs}]$ & const  \\ \hline \hline
       \multirow{2}{*}{Coefficients} & $ 0.590$ & $0.278$ & $1$ & $-0.761$ & $0.347$ \\
        & $\pm 0.092$ & $\pm 0.060$ & - & $\pm 0.060$ & $\pm 0.018$ \\
       Main-text  & $0.680$ & $0.206$ & $1$ & $-0.815$ & $0.346$ \\
       Deviation [$\sigma$]  & $0.98$ & $-1.2$ & - & $-0.9$ & $-0.06$
    \end{tabular}
    \caption{\figcap{Robustness of the quadratic XY decision boundary to $X$-$Z$ snapshot pairing.} Mean and standard deviation of the normalized coefficients obtained from ten independently re-paired and retrained networks, compared with the main-text result in \eq~\eqref{eq:decision_boundary_in_correlators}. All main-text coefficients lie within $1.2\sigma$ of the re-pairing ensemble.}
    \label{tab:XZ_random_repairings}
\end{table}

The resulting bottleneck representations are highly stable, as presented in \tab~\ref{tab:XZ_random_repairings}. Across all ten re-pairings, TetrisCNN selects the same three branches, whose leading Boolean-Fourier components correspond to $C^X[\bs]$, $C_{\mathrm{rot}}^Z[\bs\bs]$, and $C_{\mathrm{rot}}^Z[\sus{\bs\square\\\square\bs}]$. Moreover, the fitted quadratic decision boundary has the same functional form across runs. As shown in \tab~\ref{tab:XZ_random_repairings}, after normalization, all coefficients of the main-text decision boundary lie within $1.2$ standard deviations of their means across the ten re-pairings, which demonstrates that both the selected physical correlators and the decision boundaries reported in the main text are robust to the arbitrary pairing of $X$- and $Z$-basis snapshots.

\begin{figure}[t]
    \centering
    \includegraphics[width=0.98\columnwidth]{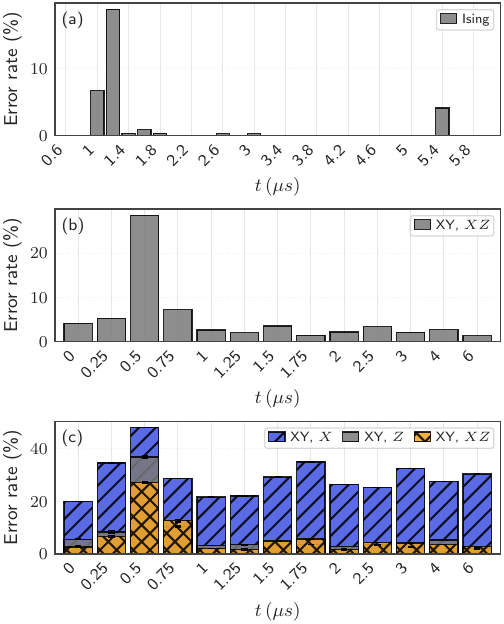}
    \caption{\figcap{Classification validation error rates per time for TetrisCNN} (a) from the main text, trained on the Ising dataset, (b) from the main text, trained on the XY dataset, and (c) trained on the XY dataset when using only $X$-snapshots, only $Z$-snapshots, and both $XZ$ snapshots in the comparable setup with the same total number of snapshots. All networks make their biggest mistakes around the detected crossover (Ising) or phase transition (XY), partly because single-snapshot classification limits accuracy. In the XY dataset in panel (c), the network trained on $Z$ snapshots only has better accuracy than the one trained on $X$ snapshots only ($91.19$\% and $70.28$\%, respectively). Information from the $X$ basis helps mostly around the phase transition, allowing the network trained on both bases to increase its total validation accuracy to $93.07$\%.} \label{fig:X_Z_XZ_errors}
\end{figure}

\begin{figure*}[t]
    \centering
    \includegraphics[width=0.98\textwidth]{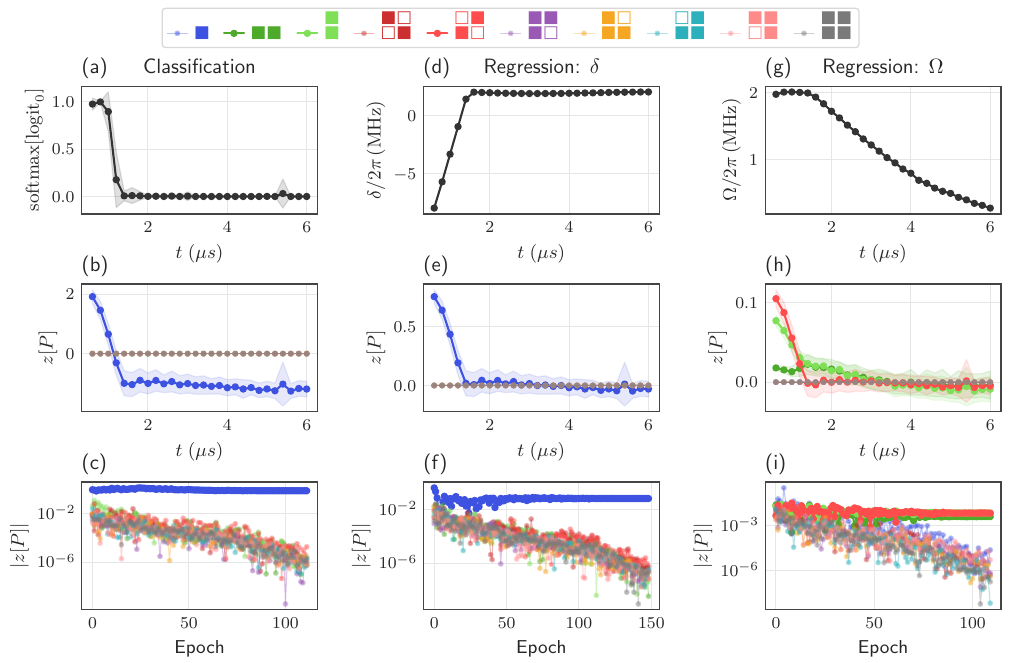}
    \caption{\figcap{Task dependence of the learned TetrisCNN representation.} Comparison of classification (a--c), regression of $\longitudinal$ (d--f), and regression of $\transverse$ (g--i) for the Ising dataset, using the same TetrisCNN architecture and branch regularization. Top row: classification probability or regression target versus sweep time. Middle row: branch activations $z[P]$ versus sweep time. Bottom row: mean absolute branch activations during training.}
    \label{fig:task_dependence}
\end{figure*}

\subsection{Contribution of the \texorpdfstring{$X$}{X} and \texorpdfstring{$Z$}{Z} measurement bases to classification}\label{app:separate_X_Z_training}

To quantify the information available in the two measurement bases, we construct XY datasets containing the same total number of experimental snapshots, $n=14471$, sampled at the same sweep points: $n$ $X$-basis snapshots, $n$ $Z$-basis snapshots, and $n/2$ snapshots from each basis in the combined $XZ$ dataset. We train TetrisCNN independently on each dataset using five random initializations. As shown in \fig~\ref{fig:X_Z_XZ_errors}, the $Z$-basis snapshots alone contain most of the information required for classification, reaching $91.19\%$ validation accuracy, compared with $70.28\%$ for the $X$ basis alone. Combining the two bases further increases the accuracy to $93.07\%$ (with standard deviations across five random initializations below $0.01\%$ for all three networks), despite using only half as many paired inputs, with the largest improvement occurring near the identified phase transition. This suggests that the $X$-basis snapshots contain complementary information about the transition, consistent with their sensitivity to the $XY$ order parameter discussed in the main text.

To understand what information is extracted from each basis, we additionally fit decision boundaries to the TetrisCNNs trained on the $X$- and $Z$-basis snapshots separately. Across all five initializations, the $X$-only classifier learns a quadratic boundary in the $X$ magnetization,
\begin{equation}
    (C^X[\bs])^2 + 0.85\,C^X[\bs] - 0.36 = 0,
\end{equation}
consistent with the relation between $(C^X[\bs])^2$ and the $XY$ order parameter discussed in the main text. However, its substantially lower classification accuracy shows that this information alone is insufficient to reliably distinguish the two classes from individual snapshots. In contrast, the $Z$-only classifier learns the stable linear boundary
\begin{equation}
    0.776\,C_{\mathrm{rot}}^Z[\bs\bs]
    - C_{\mathrm{rot}}^Z[\sus{\bs\square\\\square\bs}]
    + 0.046 = 0,
\end{equation}
involving the same two $Z$-spin correlators that dominate the combined $XZ$ classifier. Thus, TetrisCNN supplements the physically expected information from the squared $X$-magnetization with $Z$-basis correlators that provide substantially stronger discrimination at the level of individual snapshots.

\subsection{Task dependence of the learned descriptors}\label{app:task_dependence}

The main-text Discussion argues that the correlators selected by TetrisCNN depend on the learning objective. Here we demonstrate this directly, for the Ising data and rotationally unconstrained network, by comparing the branches selected for classification with those selected by the regression (implemented within the prediction-divergence method, cf.~\app~\ref{app:pdm}) of the two experimental tuning parameters separately. \Fig~\ref{fig:task_dependence} shows classification (first column, as in the main text), regression of $\longitudinal$ (second column), and regression of $\transverse$ (third column), using otherwise equivalent TetrisCNN architectures and regularization ($\lambda_{\max} = 10^3$).

Classification and regression of $\longitudinal$ both predominantly select the single-site branch $z[\bs]$. In contrast, when regressing $\transverse$, $z[\bs]$ is suppressed and the network instead relies on $z[\bs \bs]$, $z[\sus{\bs \\ \bs }]$ and $z[\sus{\square \bs \\ \bs \square}]$ branches. The representation selected by the network therefore reflects not only the underlying data but also which information is required by the learning objective. In particular, classification with TetrisCNN favors the simplest correlators sufficient to distinguish the two classes, whereas regression can favor different correlators informative of continuous variation of the target along the experimental sweep.

\subsection{Complete mapping from filters to correlators}\label{app:full_map_to_corrs}

\begin{figure*}[ht]
    \centering
    \includegraphics[width=0.98\textwidth]{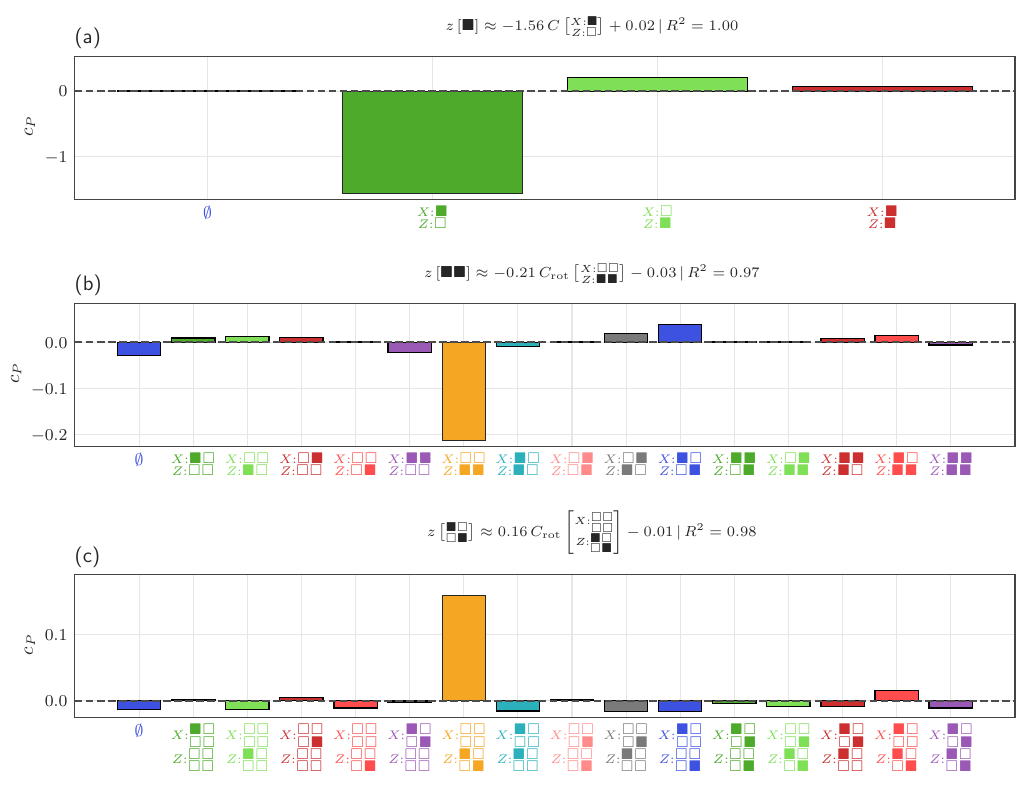}
    \caption{\figcap{Mapping from filters to correlators} via multilinear regression for three active branches in the XY model, enabled by the Boolean Fourier expansion, \eq~\eqref{eq:sum_of_correlators}. We only consider correlators $C[P'], P'\subseteq P$ defined by patterns $P'$ that are contained in the filter pattern $P$ (see \tab~\ref{tab:Tetris_correlators}). Note that we need to take into account both $C[\sus{\bs \\ \sq}]$ and $C[\sus{\sq \\ \bs}]$ because edge effects make these two correlators have different values. This is the full version of the Fig.~\ref{fig:XY_multilinear_regression} in the main text.}
    \label{fig:XY_multilinear_regression_full}
\end{figure*}

Figure~\ref{fig:XY_multilinear_regression_full} presented in this appendix reproduces \fig~\ref{fig:XY_multilinear_regression} of the main text without pruning: every subpattern correlator $C[P'], P'\subseteq P$ enters the multilinear fit, whereas the main-text version collapses the smaller positive and negative contributions into the two ``other'' categories for readability. The two fits agree by construction, since the main-text figure is obtained from this full decomposition by pruning; it is reproduced here in full for transparency.

\section{\texorpdfstring{$R$}{R}-squared for multivariate and aggregate output}\label{app:multivariate_R2}
When assessing the quality of a regression model where the predicted output is multivariate (e.g. when predicting both $\delta$ and $\Omega$ for the TFI model in the main text), common libraries (\texttt{PyTorch} \cite{PyTorch2} and \texttt{scikit-learn} \cite{scikit-learn}) allow a choice between ``uniform weighting'' and ``variance weighting'' the $R^2$ of the two outputs. We argue here that variance weighting is appropriate if the canonical interpretation of $R^2$ is to be preserved (i.e. as a proportion of variance in the data explained by the model).

That is, for multivariate $y\in\mathbb{R}^{K}$, with $N$ datapoints for each $k\in[K]$, $R^2$ for the model as a whole is defined as:
\begin{equation}
\begin{aligned}
R_{\text {tot }}^2&=1-\frac{S S_{\text {res }}}{S S_{\text {tot }}}=1-\frac{\sum_k \sum_{n \in \Gamma_k}\left(\hat{y}_k^{(n)}-y_k^{(n)}\right)^2}{\sum_k \sum_{n \in \Gamma_k}\left(\hat{y}_k^{(n)}-\bar{y}_k\right)^2} \label{eq:Rsquared_tot}
\end{aligned}
\end{equation}
where $\bar{y}_k=\frac{1}{N_k}\sum_{n\in\Gamma_k} y_n$ is the mean for group $k$. We use the subscript ``tot'' to distinguish from $R^2$ for an individual output $y_k$:
$$
\begin{aligned}
R_k^2=1-\frac{S S_{\text {res }}^k}{S S_{\text {tot }}^k}&=1-\frac{\sum_{n \in \Gamma_k}\left(\hat{y}_k^{(n)}-y_k^{(n)}\right)^2}{\sum_{n \in \Gamma_k}\left(y_k^{(n)}-\overline{y}\right)^2} 
\end{aligned}
$$
Using $S S_{\text {res }}^k=S S_{\text {tot }}^k\left(1-R_k^2\right)$ one finds after some algebra that:
$$
\begin{aligned}
R_{\text {tot }}^2=1-\sum_k \frac{S S_{\text {res }}^k}{\sum_{k'} S S_{\text {tot }}^{k'}}=\sum_k \frac{S S_{\text {tot }}^k}{S S_{\text {tot }}} R_k^2.
\end{aligned}
$$
I.e. one obtains a `variance-weighted' average of the individual $R^2_k$. I.e. it is clear that the ``average'' $R^2=\frac{1}{K} \sum_k R_k^2$ is in general different from the ``total'' $R^2$, defined by \eqref{eq:Rsquared_tot}:
$$
\frac{1}{K} \sum_k R_k^2 \neq R_{\text {tot }}^2.
$$ 

The same distinction between individual and grouped datapoints arises for the aggregated $R^2$ tracked alongside the MSE loss in the prediction-divergence method (\app~\ref{app:pdm}),
\begin{equation}\label{eq:R2}
R_{\text{agg}}^2=1-\frac{\sum_{y} \left(y-\langle \hat{y}\rangle_y \right)^2}{\sum_y\left(y-\Tilde{y}\right)^2}\,,
\end{equation}
where the sums run over unique values of $y$ and $\Tilde{y}=\frac{1}{\datasize}\sum_y y$ is their mean. $R_{\text{agg}}^2$ quantifies the variance in $y$ explained by \textit{groups} of snapshots sharing a label, whereas the standard $R^2$ measures the variance explained by individual datapoints. For multivariate $y$ it is variance weighted, following the argument above.

\section{Fitting the decision boundary to the classifier predictions in the bottleneck space}\label{app:fitting_decision_boundary}
\subsection{Fitting the decision boundary}\label{app:dec_bound_log_reg_fit}

For the XY dataset, where the TetrisCNN decision boundary is a surface in the 3D bottleneck space rather than a single threshold as for Ising, extracting an analytical approximation requires choosing its functional form and the terms included in it. To make these choices systematic, we construct a logistic-regression surrogate of the trained TetrisCNN classifier in its interpretable bottleneck space. This approach is conceptually related to surrogate-model explainability methods such as LIME~\cite{Ribeiro2016LIME}, but here we fit the surrogate globally rather than locally and use it to extract an analytical approximation of the decision boundary in an already interpretable latent space.

For each snapshot $S$, the surrogate takes the three active bottleneck
coordinates
\begin{equation}
    \mathbf{z}(S)=
    \left(
    z[\bs],\,
    z[\bs\bs],\,
    z[\sus{\bs\square\\\square\bs}]
    \right)
\end{equation}
and their polynomial combinations up to degree two, while its target is the class $\hat y_{\mathrm{Tetris}}(S)$ predicted by TetrisCNN rather than the ground-truth label. Defining
\begin{equation}
    \boldsymbol{\phi}(\mathbf z)
    =
    (z_1,z_2,z_3,z_1^2,z_2^2,z_3^2,z_1z_2,z_1z_3,z_2z_3),
\end{equation}
we fit the logistic-regression surrogate
\begin{equation}
    p(\hat y_{\mathrm{Tetris}}=1\mid\mathbf z)
    =
    \sigma\!\left[
    \beta_0+\boldsymbol{\beta}^{\mathsf T}\boldsymbol{\phi}(\mathbf z)
    \right]
    \equiv \sigma[F(\mathbf z)],
    \label{eq:logreg_surrogate}
\end{equation}
where $\sigma(x)=\frac{1}{1+e^{-x}}$ is the sigmoid function. Since $\sigma(0)=1/2$, the analytical approximation to the decision boundary
is the zero-level set
\begin{equation}
    F(\mathbf z)
    =
    \beta_0+\boldsymbol{\beta}^{\mathsf T}\boldsymbol{\phi}(\mathbf z)
    =0.
\end{equation}
Accordingly, throughout the main text and below, we report the polynomial decision function $F(\mathbf z)$ rather than the logistic probability itself. All model selection is performed using only the training snapshots, with five-fold stratified cross-validation. The reported accuracy is evaluated on the held-out validation snapshots and measures agreement with the TetrisCNN predictions.

\begin{figure}[t]
    \centering
    \includegraphics[width=0.98\linewidth]{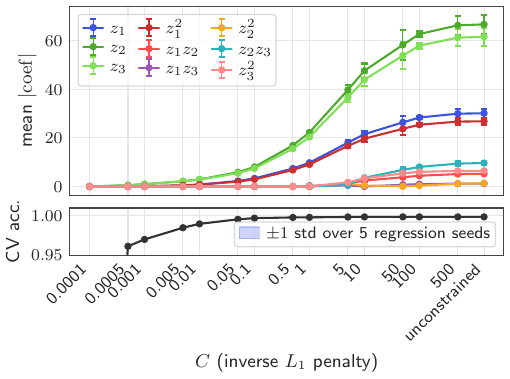}
    \caption{\figcap{Regularization path for the polynomial surrogate of the decision boundary of a single TetrisCNN.} Coefficients of the polynomial features are shown as a function of the inverse $L_1$ regularization strength $C$, together with the five-fold cross-validation accuracy. Four terms dominate as the regularization is relaxed: $z[\bs]^2$, $z[\bs]$, $z[\bs\bs]$, and $z[\sus{\bs\square\\\square\bs}]$. Error bars show the standard deviation over five cross-validation realizations.}
    \label{fig:logreg_regpath}
\end{figure}

\begin{figure}[t]
    \centering
    \includegraphics[width=0.98\linewidth]{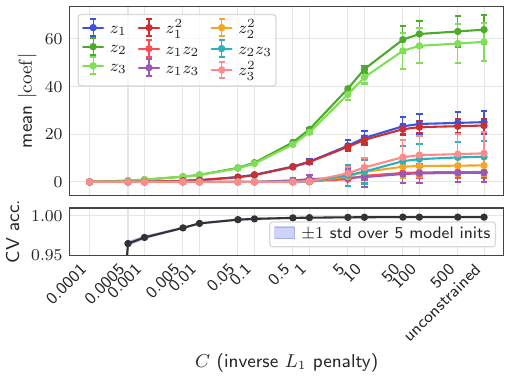}
    \caption{\figcap{Stability of the regularization path across several TetrisCNN run initializations.} Regularization paths averaged over five independently initialized TetrisCNN models. The same four dominant polynomial terms are consistently recovered, demonstrating that the inferred functional form of the decision boundary is stable across TetrisCNN random initializations.}
    \label{fig:logreg_regpath_averaged}
\end{figure}

To identify the polynomial terms most relevant to the decision boundary, we examine an $L_1$-regularized logistic-regression path, varying the inverse regularization strength $C$ from strong regularization toward the effectively unregularized limit. As shown in \fig~\ref{fig:logreg_regpath}, four terms dominate: $z[\bs]^2$, $z[\bs]$, $z[\bs\bs]$, and $z[\sus{\bs\square\\\square\bs}]$. Through the Boolean-Fourier mapping (\fig~\ref{fig:XY_multilinear_regression}), these correspond predominantly to $(C^X[\bs])^2$, $C^X[\bs]$, $C_{\mathrm{rot}}^Z[\bs\bs]$, and $C_{\mathrm{rot}}^Z[\sus{\bs\square\\\square\bs}]$, respectively. The same four terms are consistently recovered across independent TetrisCNN initializations (\fig~\ref{fig:logreg_regpath_averaged}), motivating the linear and quadratic approximations considered below. Further fitting details are given in \tab~\ref{tab:logreg_hyperparams}.

\begin{table}[t]
\centering
\caption{\figcap{Logistic-regression settings used to fit the polynomial
decision-boundary approximations.}}
\label{tab:logreg_hyperparams}
\begin{tabular}{@{}ll@{}}
\toprule
\textbf{Setting} & \textbf{Value} \\
\midrule
Target & TetrisCNN predicted class \\
Preprocessing & \texttt{StandardScaler} \\
Cross-validation & 5-fold stratified \\
Final regularization & $L_2$, $C=10^{4}$ \\
Solver & \texttt{lbfgs} \\
Class weight & balanced \\
Max. iterations & $2\times10^{4}$ \\
Evaluation & held-out validation set \\
\bottomrule
\end{tabular}
\end{table}

\subsection{Decision boundary: plane vs quadratic surface}\label{app:plane_vs_surface}

Having established the dominant terms, we next ask how much of the TetrisCNN decision rule progressively more expressive physical approximations can capture.

Remarkably, the linear approximation using two $Z$-correlators alone already reproduces the classifier predictions with $96.53\%$ accuracy:
\begin{equation}
C_{\mathrm{rot}}^Z [\bs \bs] - 0.879 C_{\mathrm{rot}}^Z [\sus{\bs\square \\ \square \bs}] + 0.41 = 0 \,.
    \label{eq:linear_only_Z}
\end{equation}
Thus, the majority of the classifier's decision rule can be explained without the $X$-magnetization contribution. 

Including $C^X[\bs]$, the full linear approximation becomes
\begin{equation}\label{eq:XY_linear_boundary}
0.22 C^{X}[\bs] + C^{Z}_{\mathrm{rot}}[\bs\bs] - 0.8 C^{Z}_{\mathrm{rot}}[\sus{\bs\square \\ \square \bs}] + 0.39 = 0\,,
\end{equation}
and reproduces the classifier predictions with $96.85\%$ accuracy.

\begin{figure}[t]
    \centering
\includegraphics[width=0.98\columnwidth]{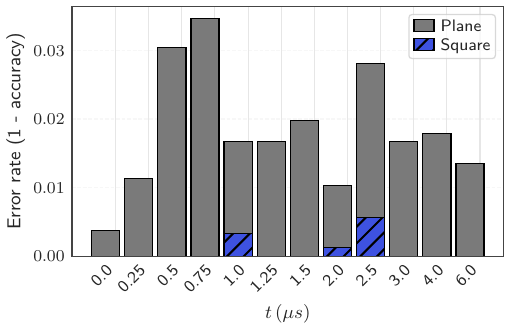}
    \caption{\figcap{Linear and quadratic approximations to the XY decision boundary.} Disagreement with the TetrisCNN predictions as a function of sweep time for the linear and quadratic approximations to the decision boundary in \fig~\ref{fig:decision_boundaries}(b). The improvement obtained by including $(C^X[\bs])^2$ is concentrated near the identified transition at approximately $0.625\,\mu\mathrm{s}$.}
    \label{fig:decision_boundary_plane_square_misclassification}
\end{figure}

Finally, only allowing a quadratic dependence on $C^X[\bs]$ as in the main text in \eq~\eqref{eq:decision_boundary_in_correlators}, improves the agreement more significantly. While the overall improvement seems modest (accuracy increases to $98.30\%$), it is concentrated near the identified phase transition, where the quadratic approximation reduces the disagreement with the classifier from approximately $3\%$ for the linear boundary to zero, as shown in \fig~\ref{fig:decision_boundary_plane_square_misclassification}. 

This shows that the two $Z$-correlators provide the dominant classification signal, while the nonlinear contribution from $C^X[\bs]$ refines the decision boundary, particularly in the vicinity of the phase transition.

\section{Limitations of symbolic regression}\label{app:SR}
In the original TetrisCNN pipeline \cite{cybinski2024tetriscnn}, we used symbolic regression to obtain an analytical approximation of the task network acting on the interpretable bottleneck. Here we revisit this approach and find that fitting the full task-network output can yield symbolic expressions that agree well with network predictions in distribution but extrapolate poorly.

\subsection{Search space and evaluation}\label{app:sr_search_space}

We use \texttt{PySR}~\cite{cranmer2023pysr}, which searches over symbolic expressions using an evolutionary algorithm. We restrict the operator basis to
\begin{equation}
    \{+,\,\times,\,\mathrm{negation},\,(\cdot)^2,\,(\cdot)^3,\,(\cdot)^4\},
\end{equation}
favoring low-order polynomial forms natural for local spin observables. Multiplication is restricted to multiplication by a constant, powers cannot be nested, and expression complexity is explicitly penalized. We use mean absolute error as the fitting loss and a parsimony penalty of $\lambda=0.01$. Expanding the operator basis, for example by including division or exponentials, does not qualitatively change the conclusions below.

Each symbolic expression is fitted to reproduce the output of the trained task network from the active bottleneck variables: the classification logit for the Ising classifier or the predicted transverse field $\transverse$ for the Ising regression task. In-distribution agreement is quantified by $R^2_{\mathrm{val}}$ on validation snapshots.

To test whether the symbolic form captures the task network beyond the training distribution, we evaluate it on synthetic spin configurations not encountered during training. Because the targets for these configurations are provided by the output of the trained network itself, we are free to probe arbitrary spin configurations. We consider paramagnetic, ferromagnetic, and antiferromagnetic families, each generated from a fixed representative configuration with four random spin flips. These configurations are passed through the original network and the symbolic surrogate, and the agreement is evaluated separately for each family. Alongside $R^2$, we report the Spearman correlation $\rho$, which distinguishes quantitatively poor fits that nevertheless preserve the ordering of the network predictions.

\begin{figure*}[t]
\centering
\includegraphics[width=\linewidth]{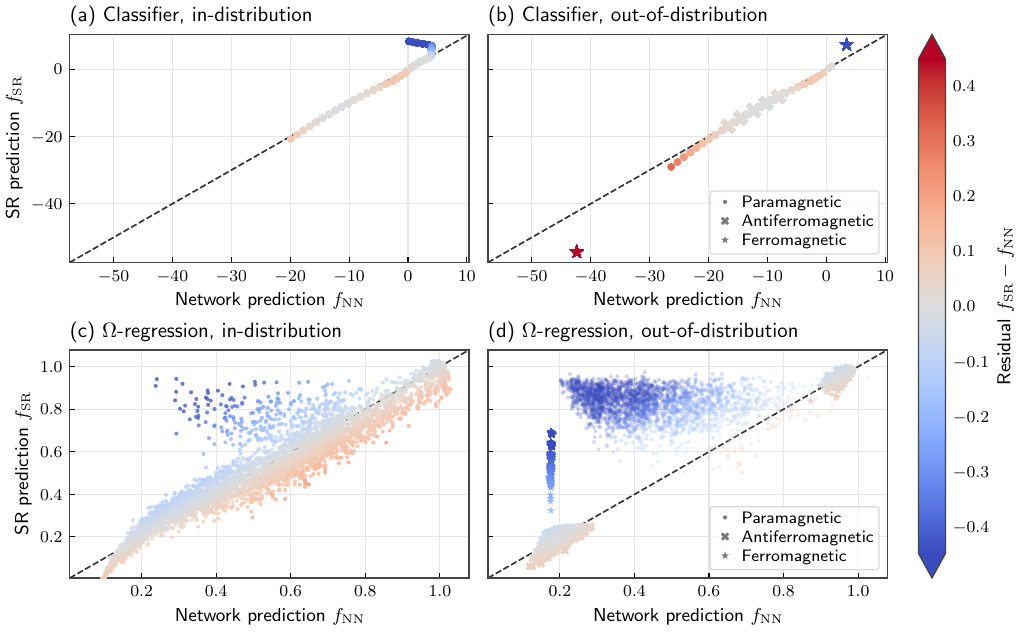}
\caption{\figcap{In-distribution agreement does not guarantee out-of-distribution robustness.}
Symbolic-regression prediction $f_{\mathrm{SR}}$ against network prediction $f_{\mathrm{NN}}$ for the equations marked $\dagger$ in \tab~\ref{tab:sr_ood}. Top: Ising classifier; bottom: Ising $\transverse$ regression. Left panels show validation data and right panels the paramagnetic, ferromagnetic, and antiferromagnetic probe families. Points are colored by the signed residual $f_{\mathrm{SR}}-f_{\mathrm{NN}}$.}
\label{fig:sr_parity_ood}
\end{figure*}

\subsection{Classification: full network output versus decision boundary}\label{app:sr_classification}

The Ising classifier provides the simplest setting, since its bottleneck collapses to the single branch $z[\bs]$. Symbolic regression is trained to reproduce the classifier logit over the full bottleneck domain. Several expressions on the PySR Pareto front achieve essentially identical classification accuracy, while their agreement with the logit ranges from $R^2_{\rm val}=0.50$ to $1.00$ and their out-of-distribution behavior differs substantially, with the strongest degradation occurring for the ferromagnetic probe family ($R^2_{\rm val}=0.56$--$0.85$).

Importantly, this is a stricter requirement than reproducing the classifier itself, as we do in the main text and describe in detail in \app~\ref{app:dec_bound_log_reg_fit}. A classification surrogate only needs to recover the location of the decision boundary, and the detailed variation of the logit away from that boundary is irrelevant to the predicted class. We therefore evaluate the polynomial decision-boundary approximation used in the main text on the same out-of-distribution configurations. It reproduces the TetrisCNN class predictions with $100\%$ accuracy on all three out-of-distribution probe families.

The decision-boundary fit generalizes perfectly to the out-of-distribution probes, whereas the symbolic-regression fit to the full network logit does not. This may be because reproducing the full logit promotes more complicated, less generalizable functions, or because symbolic regression additionally depends on the user-defined operator grammar and complexity prior. What is clear, however, is that a controlled approximation of the decision boundary provides the more robust surrogate in this setting.

\begin{table*}[t]
\centering
\caption{\figcap{Symbolic-regression fits and out-of-distribution
generalization.}
For each equation, we report the validation fit to the network output,
$R^2_{\mathrm{val}}$, and the coefficient of determination $R^2$ and Spearman correlation $\rho$ on paramagnetic, ferromagnetic, and antiferromagnetic probe families. Rows within each task correspond to expressions of increasing complexity along a single PySR Pareto front. $\dagger$ marks the expressions shown in \fig~\ref{fig:sr_parity_ood}.}
\label{tab:sr_ood}
\resizebox{\textwidth}{!}{%
\begin{tabular}{llccccccc}
\toprule
& & & \multicolumn{2}{c}{\textbf{paramagnetic}}
& \multicolumn{2}{c}{\textbf{ferromagnetic}}
& \multicolumn{2}{c}{\textbf{antiferromagnetic}} \\
\cmidrule(lr){4-5}\cmidrule(lr){6-7}\cmidrule(lr){8-9}
\textbf{Task} & \textbf{Equation} & $R^2_\mathrm{val}$
& $R^2$ & $\rho$ & $R^2$ & $\rho$ & $R^2$ & $\rho$ \\
\midrule
Ising clf.\ & $8.60\,z[\bs]$ & $0.50$ & $1.00$ & $1.00$ & $0.84$ & $0.95$ & $1.00$ & $0.96$ \\
Ising clf.\ & $-z[\bs]^2 + 6.35\,z[\bs] - 1.22$ $^\dagger$ & $0.95$ & $0.99$ & $1.00$ & $0.85$ & $0.95$ & $1.00$ & $0.96$ \\
Ising clf.\ & $-1.55\,z[\bs]^2 + 5.69\,z[\bs] - 1.19$ & $0.99$ & $0.96$ & $1.00$ & $0.56$ & $0.95$ & $0.97$ & $0.96$ \\
Ising clf.\ & $-0.419\,z[\bs]^3 - 1.73\,z[\bs]^2 + 6.73\,z[\bs] - 0.481$ & $1.00$ & $1.00$ & $1.00$ & $0.76$ & $0.95$ & $1.00$ & $0.96$ \\
\midrule
Ising reg.\ ($\transverse$) & $11.5\,z\qty[\sus{\bs\\\bs}] + 0.386$ & $0.82$ & $-5.9$ & $-0.31$ & $-4.8$ & $-0.90$ & $-0.62$ & $0.72$ \\
Ising reg.\ ($\transverse$) & $-1.11\times10^{3}\,z\qty[\sus{\bs\\\bs}]^3 + 13.8\,z\qty[\sus{\bs\\\bs}] + 0.407$ & $0.91$ & $-7.1$ & $-0.31$ & $-0.04$ & $0.90$ & $-0.26$ & $0.72$ \\
Ising reg.\ ($\transverse$) & $z\qty[\bs\bs] - 1.22\times10^{3}\,z\qty[\sus{\bs\\\bs}]^3 + 12.9\,z\qty[\sus{\bs\\\bs}] + z\qty[\sus{\sq\bs\\\bs\sq}] + 0.406$ $^\dagger$ & $0.92$ & $-6.3$ & $-0.25$ & $0.38$ & $0.87$ & $0.12$ & $0.76$ \\
\bottomrule
\end{tabular}%
}
\end{table*}

\subsection{Higher-dimensional bottleneck: failure to extrapolate}\label{app:sr_richer_bottleneck}

The limitations become clearer for the Ising $\transverse$-regression task. Here the single-site branch is suppressed
(\app~\ref{app:task_dependence}), and the learned representation is distributed across several two-body branches. No expression on the PySR Pareto front is simultaneously simple, accurate in distribution, and robust across the out-of-distribution probes.

In particular, $R^2_{\mathrm{para}}$ is negative for every expression in \tab~\ref{tab:sr_ood}, and the simpler candidates also fail on the ferromagnetic and antiferromagnetic families. The Spearman correlation shows that these failures can differ qualitatively: for example, one expression reverses the ferromagnetic ordering ($\rho_{\mathrm{ferro}}=-0.90$), whereas another preserves it ($\rho_{\mathrm{ferro}}=0.90$) despite still having poor $R^2$.

The out-of-distribution predictions are not arbitrary [\fig~\ref{fig:sr_parity_ood}(d)]: they remain structured by configuration family and largely within the numerical range encountered in distribution. Nevertheless, their quantitative disagreement with the network demonstrates that a good symbolic fit on the available validation data does not identify a unique or robust functional form.

Taken together, these results motivate a more restricted role for symbolic regression in TetrisCNN. For classification, fitting the full task-network output is unnecessarily demanding when the scientific object of interest is the decision boundary: a simple polynomial fit to that boundary generalizes perfectly across the out-of-distribution probes considered here. For regression, where no analogous boundary reduction is available, symbolic regression can fit the network well in distribution yet extrapolate poorly. Moreover, the selected symbolic form depends on the user-defined operator grammar and complexity prior. We therefore use architecture-native interpretation to identify the physical bottleneck variables, and, when possible, fit only the simplest task-relevant object in that interpretable space rather than attempting to reconstruct the entire task network symbolically.

\end{document}